\documentclass[amsmath,amssymb,aps,pre,twocolumn,floatfix]{revtex4-2}
\usepackage{multirow}
\usepackage{graphicx}
\usepackage{color}
\usepackage{dcolumn}
\usepackage{bm}
\usepackage{hyperref}
\usepackage[mathlines]{lineno}
\usepackage{comment}
\usepackage[T1]{fontenc}
\usepackage[normalem]{ulem}

\begin{document}

\title{Wang-Landau study of lattice gases on geodesic grids}

\author{Gabriele Costa}
\author{Santi Prestipino}

\affiliation{Dipartimento di Scienze Matematiche e Informatiche, Scienze Fisiche e Scienze della Terra, Università degli Studi di Messina, Viale F. Stagno d’Alcontres 31, 98166 Messina, Italy}

\begin{abstract}
We study a family of lattice-gas systems defined on semiregular grids, obtained by projecting the vertices of three different geodesic icosahedra onto a spherical surface.
By using couplings up to third neighbors we explore various interaction patterns, ranging from core-corona repulsion to square-well attraction and short-range attractive, long-range repulsive potentials.
The relatively small number of sites in each grid ($\sim 100$) enables us to compute the exact statistical properties of the systems as a function of temperature and chemical potential by Wang-Landau sampling.
For each case considered we highlight the existence of distinct low-temperature ``phases'', featuring, among others, regular-polyhedral, cluster-crystal, and worm-like structures.
We highlight similarities and differences between these motifs and those observed on the triangular lattice under the same conditions.
Finally, we discuss the relevance of our results for the bottom-up realization of spherical templates with desired functionalities.
\end{abstract}
\maketitle

\section{Introduction}

Reconstructing the equilibrium behavior of a many-particle system is the main objective of a statistical-mechanics calculation.
This task is usually accomplished by means of Metropolis Monte Carlo simulation, a very general and powerful method which, however, becomes inefficient when the Hamiltonian contains many parameters, or close to a first-order transition point, where the sample can remain trapped in a local free-energy minimum for long.
In this respect, finding exactly solvable models for the problem at hand would be crucial, since, besides feeding our insight and knowledge, exact calculations best illustrate the characteristics of the model that are responsible for specific emergent behaviors.
In particular, solvable systems of colloidal-like particles could be helpful in discerning which features of the interaction are essential for the appearance, at low temperature, of unusual ordered phases like cluster crystals or stripes~\cite{zhuang2016recent}.
The latter morphologies are examples of mesoscale structures (``microphases'') which, in soft-matter and biological systems, are nearly as frequent as solid-liquid-vapor triarchy in simple fluids.
Indeed, microphases arise in systems as diverse as block copolymers~\cite{Bates2016,Reddy2018}, Langmuir films~\cite{Keller1986}, and protein solutions~\cite{Stradner2004}, having in common a long-range interparticle repulsion arising from screened electrostatic interactions~\cite{Sear1999,Ciach2013,Ciach2018}. 

Our study  of self-assembly in particle systems will be easier if we use a discrete embedding space, like in a lattice-gas model where particles are defined on the sites of a regular grid (usually, a Bravais lattice).
By enforcing the condition of single site occupancy, particles gain a hard core and the numerical analysis is further simplified.
Clearly, to observe non-trivial behavior the grid size cannot be too small, nor the range of interactions too short.
To add more interest to our investigation, we choose an ambient space with intrinsic curvature, e.g., a grid with the topology of the sphere, since then the particle interaction will typically be frustrated and observing a complex spatial organization in a relatively small system is more likely.
Using a closed grid has an another advantage:
we should not bother with boundary conditions to make the various orders fit into the grid.

In this paper we analyze the equilibrium behavior of classical particles on polyhedral grids (i.e., grids made of the vertices and edges of a convex polyhedron).
To keep the effects of a curved space to a minimum, our choice goes to geodesic icosahedra~\cite{visualpolyhedra}, which is a class of semiregular polyhedra with triangular faces and the least possible number (12) of disclinations.
For such systems a rich interplay can be expected between ``phases'' of a various nature --- on a finite grid, genuine thermodynamic phases only exist at zero temperature ($T=0$); at $T>0$, any phase transition is replaced by a crossover region~\cite{binder1987finite}.
At variance with Refs.~\cite{Pekalski2018,Dlamini2021}, it is not our purpose to examine a specific fluid of particles; rather, we span a whole range of interactions and for each one  the leading structures are identified.
Clearly, a key factor will be having a fast method to extract the system phases from the Hamiltonian.
That is why we employ the Wang-Landau (WL) algorithm~\cite{wang2001efficient,wang2001determining}, a sophisticated variation of the Metropolis algorithm which allows the computation of the {\em exact} density of states of the system with relatively small effort.
From that, all thermodynamic properties follow at once.
In practice, the size of the system cannot be too large, since otherwise the WL method ceases to be an opportunity;
on the other hand, if the system were too big the effects of curvature will be totally obscured, which is not what we really want.

A sample of the rich structural behavior of ``spherical'' lattice gases can be found in Ref.~\cite{Costa2025}.
In that paper, we considered lattice-gas models defined on the grid based on a pentakis icosidodecahedron, which is one of the simplest geodesic icosahedra.
In the present paper, we enlarge the inventory of amusing structures by considering systems defined on larger geodesic grids, also with the purpose to draw a comparison with the phases of triangular-lattice gases.
Upon increasing the number of sites within the geodesic family, the spatial separation between the fivefold-coordinated vertices becomes progressively larger and curvature effects gradually turn off, at least insofar as the interaction is cut off at a fixed distance.

\begin{figure*}
\includegraphics[width=4.95cm]{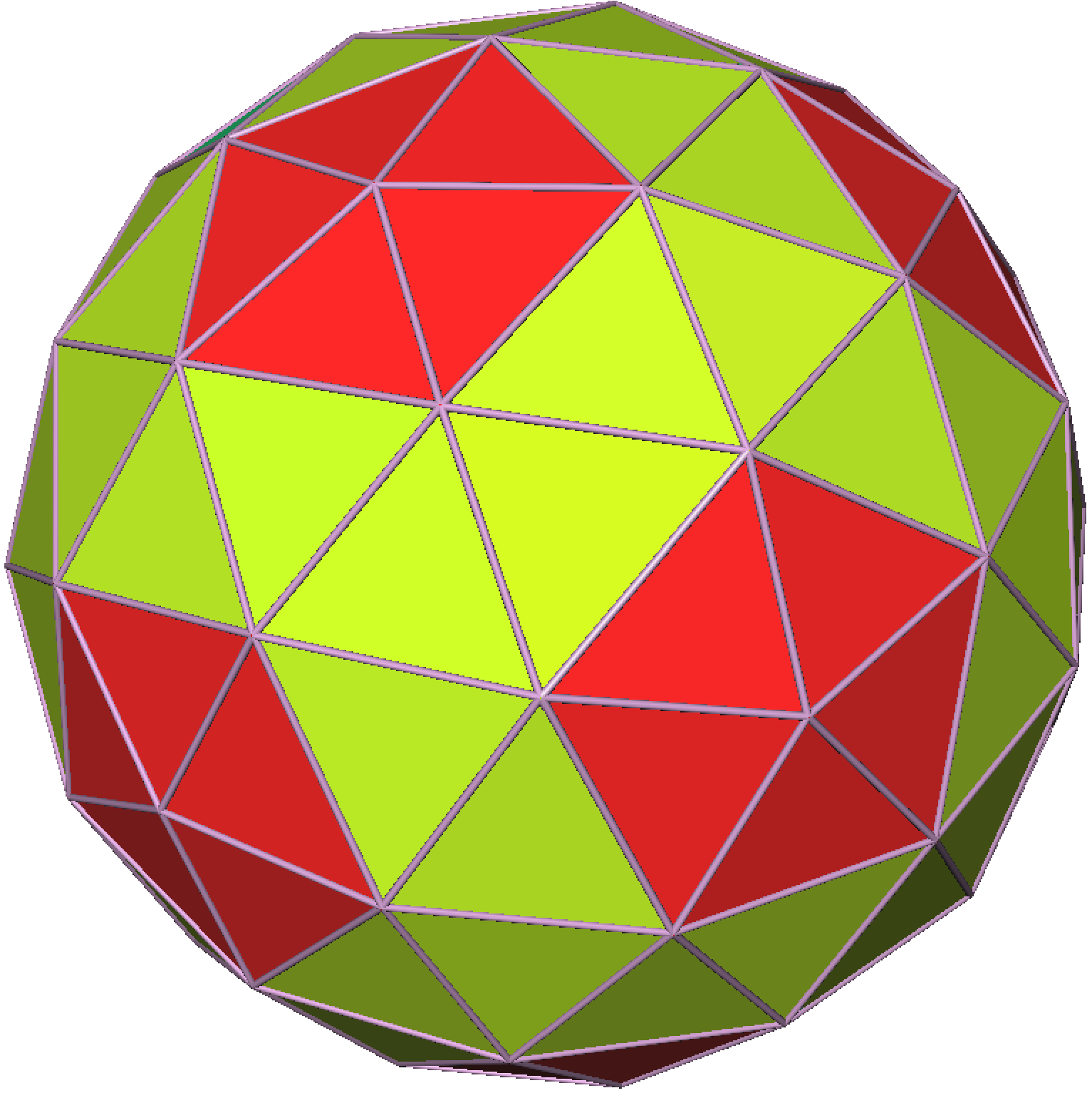}
\hfill
\includegraphics[width=5cm]{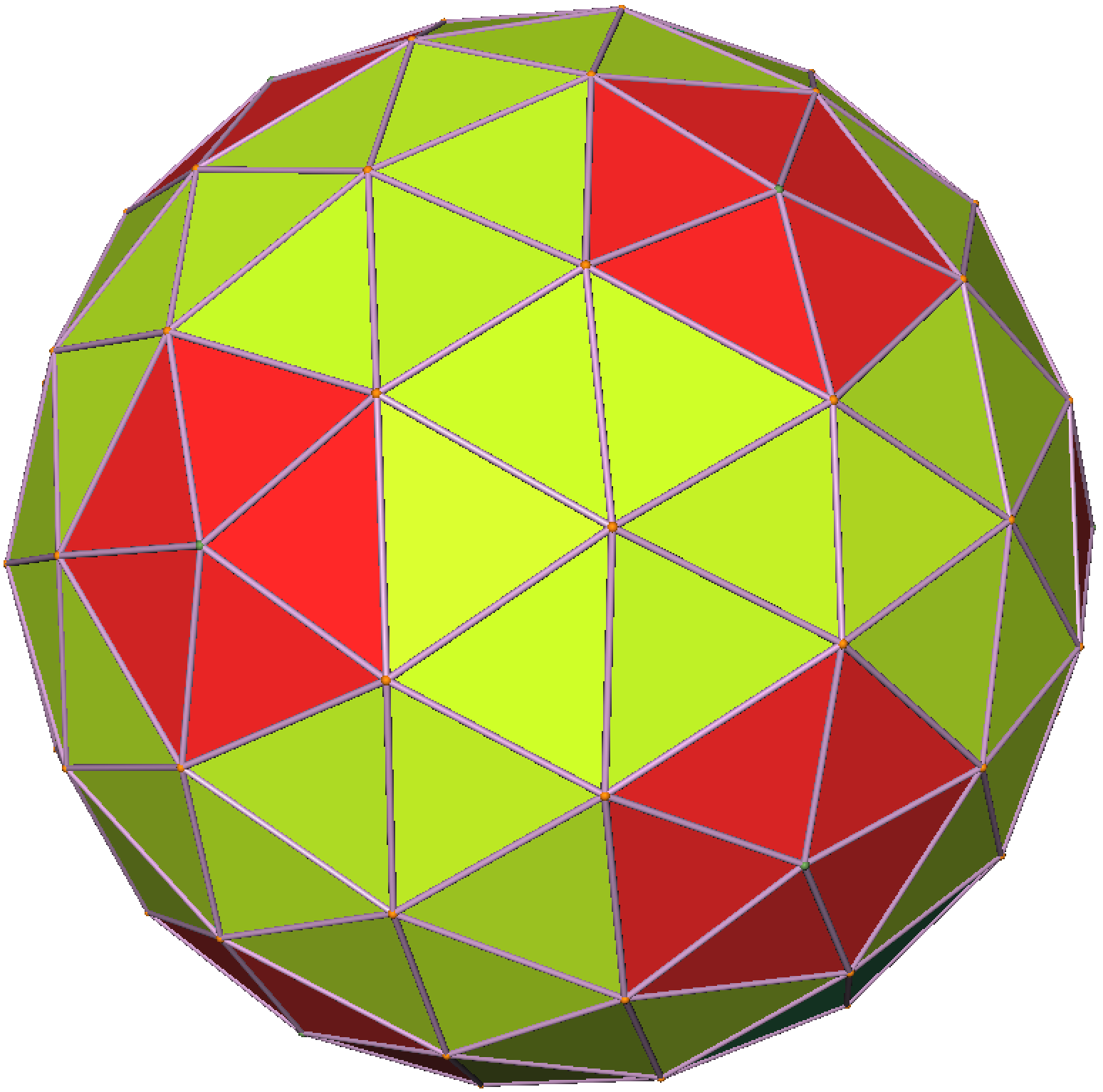}
\hfill
\includegraphics[width=5.05cm]{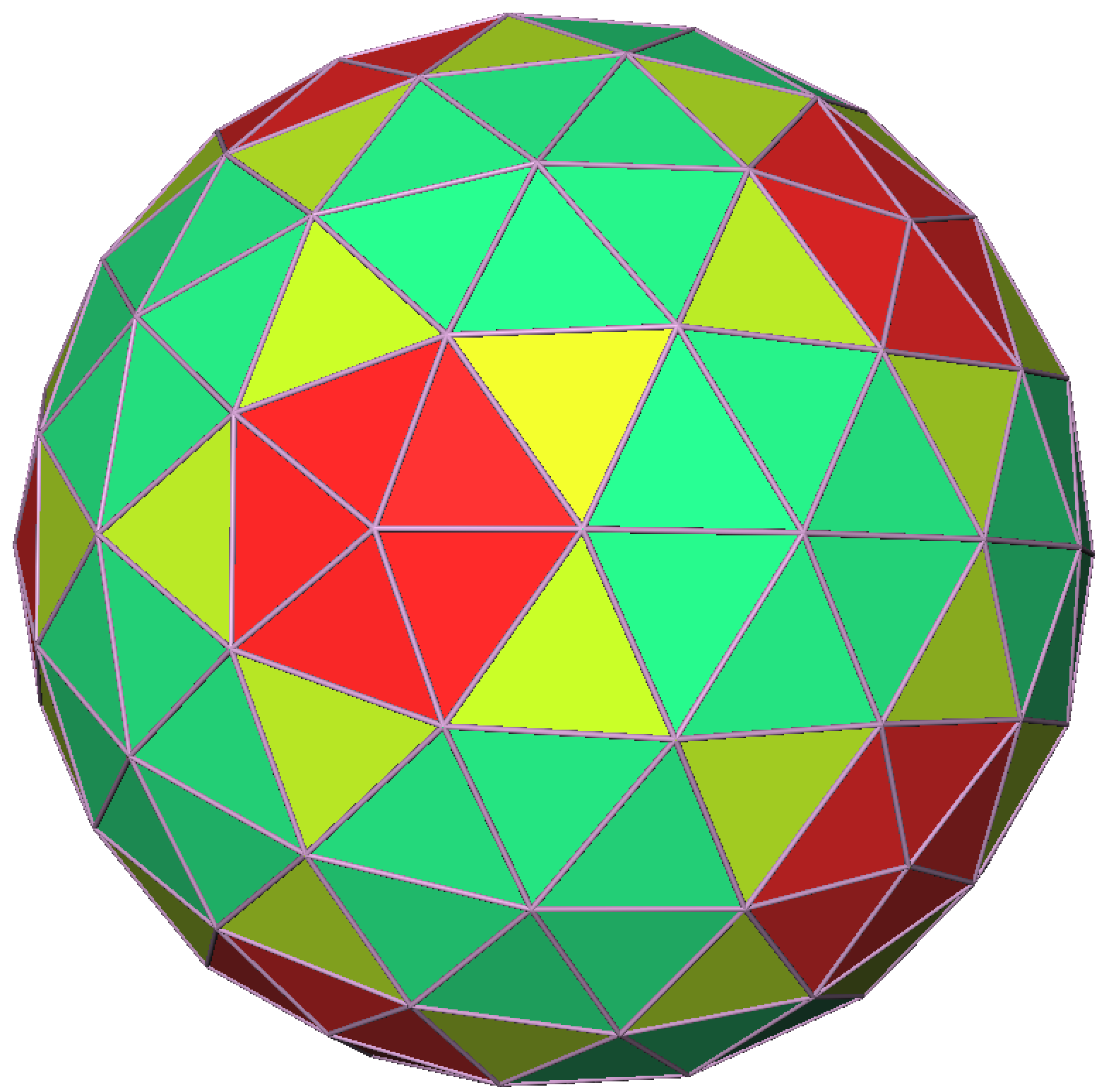}
\caption{\label{fig1}The three geodesic icosahedra considered in this paper.
From left to right: PSD, HPTI, and HPCD (see the acronyms defined in the text).
Assuming the mean first-neighbor distance as reference, the curvature of the grid gets smaller from left to right.
Each polyhedron in the figure is obtained from a corresponding parent polyhedron (e.g., the snub dodecahedron for the PSD) by raising a pyramid on each pentagonal and/or hexagonal face (Conway's kis operation).
The height of the pyramid can be chosen such that all edges are tangent to the unit sphere.
Different colors are used for different types of triangles.
The total number of vertices is 72 for PSD, 92 for HPTI, and 122 for HPCD.
The fivefold vertices are twelve in any case.
}
\end{figure*}

The plan of the paper is as follows.
In Section 2, we introduce a class of discrete particle systems and describe the method used to compute their statistical properties.
Results are presented and commented in Section 3.
Finally, Section 4 is devoted to conclusions and perspectives.

\section{Model and method}

In our language, a grid is a network made of sites and edges connecting adjacent sites (here, the notion of distance is inherited from three-dimensional Euclidean space).
Particles live on the sites;
calling $i$ (from 1 to $M$) the generic site and $c_i$ (either 0 or 1) its occupation number, the Hamiltonian of a lattice-gas model on the grid is a function $H[c]$, where $c=\{c_1,\ldots,c_M\}$ is the generic microstate.

Up to a convenient projection (see below), all grids focused on in this paper consist of the vertices of a geodesic polyhedron, i.e., a convex polyhedron made from triangles.
While the general setup is similar to Ref.~\cite{Costa2025}, the grids considered hereafter are finer than the single one examined in our previous paper, in the aim to draw more detailed conclusions on how $H[c]$ should be designed in order that specific types of ordering appear at low temperature.
We consider three possibilities (see Fig.~1):
The pentakis snub dodecahedron (PSD), the hexakis-pentakis truncated icosahedron (HPTI), and the hexakis-pentakis chamfered dodecahedron (HPCD)~\cite{visualpolyhedra}.

In the present paper we choose the ``canonical forms'' of these three polyhedra,  in which the vertices are positioned in space so as to ensure that all edges are tangent to the unit sphere.
Therefore, the edge-scribed radius is unity by construction.
The grid sites are then obtained by projecting the vertices of the canonical form onto the unit sphere.

\begin{table*}
\begin{ruledtabular}
\begin{tabular}{ccccc}
&\multicolumn{3}{c}{\large{PSD model}}\\ \\
\textbf{Shell}&\textbf{Sixfold vertex}
&
&
&\textbf{Fivefold vertex}\\
\hline
$1$st& 0.402836 (1), 0.463857 (5)& & &0.402836 (5)\\ \\
$2$nd& 0.726567 (1), 0.750536 (2),  & & &0.726567 (5)\\ &0.795775 (3)&&&\\ \\
$3$rd& 0.846202 (1), 0.901344 (4)& & &0.846202 (5)\\
\hline
\hline
&\multicolumn{3}{c}{\large{HPTI model}}\\ \\
\textbf{Shell}
&\textbf{Sixfold v. \#1}&\textbf{Sixfold v. \#2}
&
&\textbf{Fivefold vertex}\\
\hline
$1$st& 0.348615 (1), 0.403548 (3), & 0.412411 (6) && 0.348615
 (5)\\ &0.412411 (2)&&&\\ \\
$2$nd& 0.652955 (2), 0.698966 (4)& 0.640852 (3), 0.713644
(3) && 0.640852 (5)\\ \\
$3$rd& 0.738816 (1), 0.797534 (2), & 0.797534 (6) && 0.738816 (5)\\ & 0.807096 (2) &&&\\
\hline
\hline
&\multicolumn{3}{c}{\large{HPCD model}}\\ \\
\textbf{Shell}
&\textbf{Sixfold v. \#1}&\textbf{Sixfold v. \#2}&\textbf{Sixfold v. \#3}&\textbf{Fivefold vertex}\\
\hline
$1$st& 0.295554 (1), 0.343629 (2),& 0.352635 (4), 0.362843 (2) & 0.353846 (3), 0.362843 (3)& 0.295554
 (5)\\ &  0.352635 (3), 0.353846 (1) &&&\\ \\
$2$nd& 0.556004 (2), 0.603210 (2), & 0.546533 (2), 0.618034
(4) & 0.608634 (6) & 0.546533 (5)\\ & 0.608634 (2) &&&\\ \\
$3$rd& 0.684092 (2), 0.694222 (2), & 0.684092 (4), 0.705093 (2) & 0.640852(3), 0.713644(3) & 0.640852 (5)\\ &0.705093 (1) &&&\\
\end{tabular}
\end{ruledtabular}
\caption{\label{table1} Chord distances (after the projection on the sphere) from a central vertex to its neighbors, distinguishing a fivefold vertex from (up to three different) sixfold vertices.
Within parentheses is the number of equivalent neighbors.
}
\end{table*}

We list in Table \ref{table1} the chord distances from a central site (either fivefold or sixfold) to the closest grid sites.
Using the triangular lattice as paradigm, we gather neighbor sites in shells (e.g., regarding the six nearest sites to a sixfold site as belonging to the same, first-neighbor shell), despite the distance from the central site not being the same for all (in this way, the sites in the first shell are those connected by an edge to the central site).
Calling $u_\alpha$ (with $\alpha=1,2,3$) the strength of the interaction between the central particle and the particles in the $\alpha$-th shell, the general Hamiltonian reads:
\begin{equation}
H[c]=u_1\sum_{\rm 1NP}c_ic_j+u_2\sum_{\rm 2NP}c_kc_l+u_3\sum_{\rm 3NP}c_mc_n\,,
\label{eq1}
\end{equation}
where the first sum runs over all distinct pairs of first-neighbor (1NP) sites, and so on.
We refer to (\ref{eq1}) as the $(u_1,u_2,u_3)$ interaction.
By a suitable choice of the couplings, we can represent various kinds of interaction (e.g., core-corona repulsion, Lennard-Jones type, and SALR --- standing for ``short-range attractive, long-range repulsive''~\cite{zhuang2016recent}), contrasting them with one another in relation to thermal behavior and dominant structures. 

In particular, due to the effectively hard-core nature of the particle interaction, we use the term core-corona interaction to describe a purely repulsive interaction characterized by positive or null coupling parameters (e.g., $(1,1,0)$).
In contrast, we call Lennard-Jones-type a discrete interaction featuring an attractive well, possibly preceded by a repulsive shoulder.
Finally, we classify the interaction (\ref{eq1}) as SALR whenever a short-range attraction is followed by a longer-range repulsion.
From now on, the couplings are given in $\epsilon$ units, where $\epsilon>0$ is an arbitrary energy.
In turn, the temperature $T$ is given in units of $\epsilon/k_{\rm B}$, where $k_{\rm B}$ is Boltzmann's constant. 

Once the couplings have been fixed, we compute the statistical properties of the system in the grand-canonical ensemble by the WL algorithm~\cite{landau2004new}.
The WL method allows the computation of the exact density of states of the system, $g_{{\cal N},{\cal E}}$, here given in terms of the energy (${\cal E}$) and particle number (${\cal N}$).
It manages to do so by sampling a probability density proportional to the reciprocal of the density of states.
This makes all ${\cal E}$ and ${\cal N}$ values (favorable and less favorable) to be visited (roughly) the same number of times.
In practice, $g_{{\cal N},{\cal E}}$ is updated during the run by multiplying its entry at the current ${\cal N},{\cal E}$ values by a modification factor $f$; when the histogram of energies and particle numbers becomes approximately flat, $f$ is halved and the simulation is started again with the histogram reset to zero.
The simulation is stopped when $\ln f$ becomes smaller than $10^{-7}$.
WL sampling is insensitive to the existence of multiple free-energy minima;
therefore, it is the method of choice for the study of low-temperature systems, which usually undergo a number of discontinuous changes with increasing the pressure or the chemical potential.

From the knowledge of $g_{{\cal N},{\cal E}}$, the partition function readily follows as
\begin{equation}
\Xi(T,\mu)=\sum_{{\cal N},{\cal E}}g_{{\cal N},{\cal E}}e^{\beta\mu{\cal N}}e^{-\beta{\cal E}}\,,
\label{eq2}
\end{equation}
where $\mu$ is the chemical potential and $\beta=(k_{\rm B}T)^{-1}$.
The pressure $P$ of the system is derived from the grand potential $\Omega=-k_{\rm B}T\ln\Xi$ as $P(T,\mu)=-\Omega/M$.
In a finite system close to zero temperature, the slope of $P$ as a function of $\mu$ undergoes a rapid increase near each phase crossover, while keeping roughly constant across the phase regions.

Calling $N$ and $E$ the grand-canonical averages of ${\cal N}$ and ${\cal E}$, respectively, it immediately follows that
\begin{equation}
\left.\frac{\partial P}{\partial\mu}\right|_T=\frac{N}{M}\equiv\rho
\label{eq3}
\end{equation}
($\rho$ is the number density).
At fixed $T$, the density $\rho$ is an increasing function of $\mu$, alternating plateaus (``phases'') with steps (phase crossovers), which are sharper at lower temperature.

Two further quantities are worth computing, namely the isothermal compressibility $\kappa_T$ and the entropy per site $s$, respectively given by
\begin{equation}
\kappa_T=\frac{1}{\rho^2}\left.\frac{\partial\rho}{\partial\mu}\right|_T
\,\,\,\,\,\,{\rm and}\,\,\,\,\,\,
s(T,\mu)=-\frac{1}{M}\left.\frac{\partial\Omega}{\partial T}\right|_\mu=\left.\frac{\partial P}{\partial T}\right|_\mu\,.
\label{eq4}
\end{equation}
Explicit expressions for $\kappa_T$ and $s$ are:
\begin{equation}
\rho k_{\rm B}T\kappa_T=\frac{\langle{\cal N}^2\rangle-N^2}{N}
\label{eq5}
\end{equation}
and
\begin{equation}
s(T,\mu)=\frac{1}{MT}(E+MP-\mu N)\,.
\label{eq6}
\end{equation}

Finally, let ${\cal S}_{{\cal N},{\cal E}}=k_{\rm B}\ln g_{{\cal N},{\cal E}}$ be the microcanonical entropy.
Then,
\begin{equation}
\Xi=\sum_{{\cal N},{\cal E}}e^{-\beta({\cal E}-T{\cal S}_{{\cal N},{\cal E}}-\mu{\cal N})}\equiv\sum_{{\cal N},{\cal E}}e^{-\beta\widetilde\Omega_{{\cal N},{\cal E}}}\,,
\label{eq7}
\end{equation}
where
\begin{equation}
\widetilde\Omega_{{\cal N},{\cal E}}={\cal E}-T{\cal S}_{{\cal N},{\cal E}}-\mu{\cal N}
\label{eq8}
\end{equation}
is the {\em generalized grand potential} (GGP).
Since $\widetilde\Omega_{{\cal N},{\cal E}}={\cal O}(M)$ for $M\gg 1$, application of the saddle-point theorem leads (if $\beta$ is not too small) to
\begin{equation}
\Xi\approx\exp\left\{-\beta\widetilde\Omega_{\overline{\cal N},\overline{\cal E}}\right\}\,\,\,\,\,\,{\rm with}\,\,\,\,\,\,\left(\overline{\cal N},\overline{\cal E}\right)={\rm argmin}\,\widetilde\Omega_{{\cal N},{\cal E}}\,.
\label{eq9}
\end{equation}
From the previous $\Xi$ estimate, being the more valid the larger $M$, we obtain
\begin{eqnarray}
&&\Omega\equiv\widetilde\Omega_{\overline{\cal N},\overline{\cal E}}\approx-k_{\rm B}T\ln\Xi\,;
\nonumber \\
&&P=-\Omega/M\approx(k_{\rm B}T/M)\ln\Xi\,.
\label{eq10}
\end{eqnarray}
By the saddle-point theorem, we similarly derive $N\approx\overline{\cal N},E\approx\overline{\cal E}$, and $s\approx{\cal S}_{\overline{\cal N},\overline{\cal E}}/M$.
In particular, notice the difference between an exact evaluation of $E$ (or $N$) and the value of $\overline{\cal E}$ (or $\overline{\cal N}$):
if we choose integer couplings, then $\overline{\cal E}$ is integer as well, whereas $E$ is only approximately so. 

For fixed $(T,\mu)$, the graph of $\widetilde\Omega_{{\cal N},{\cal E}}$ is a multi-valley landscape.
The absolute GGP minimum falls at $(\overline{\cal N},\overline{\cal E})$.
This global minimum jumps from one valley to another as the control parameters $(T,\mu)$ pass over a ``transition point''.
We will illustrate this behavior in one case below.

\section{Results}

We hereafter review, one grid at a time, the phase behavior of model (\ref{eq1}) in a few relevant cases, covering all the main types of interaction.
In the following, it is by far sufficient to work with integer couplings.

\subsection{PSD model}

We start showing results for a number of representative lattice gases among those defined on the PSD grid.
We especially focus on systems whose low-temperature structures are particularly symmetric or intriguing in some respect.

\begin{figure}
\centering
\includegraphics[width=0.45\linewidth]{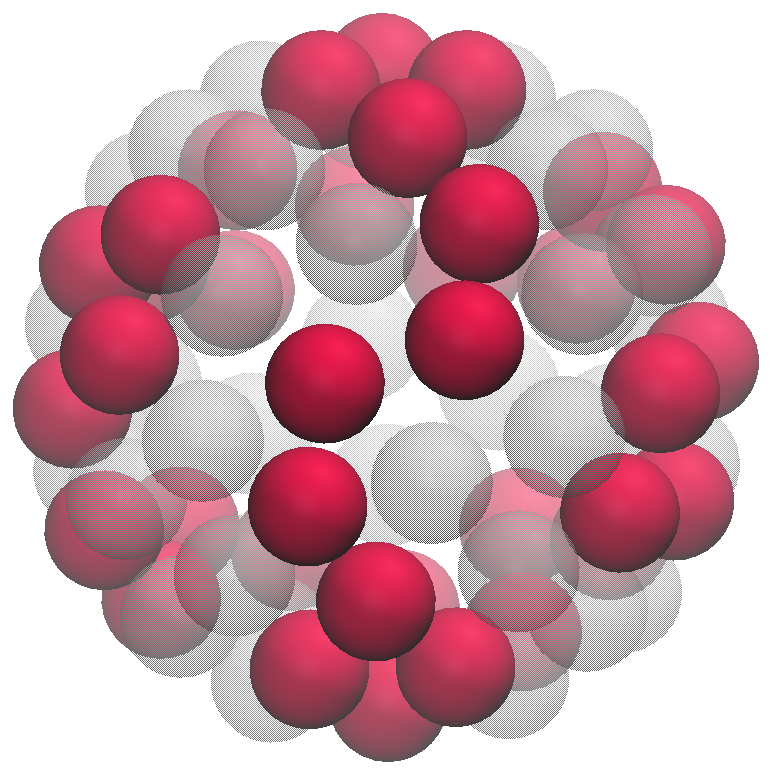}\,\,\,\,\,
\includegraphics[width=0.45\linewidth]{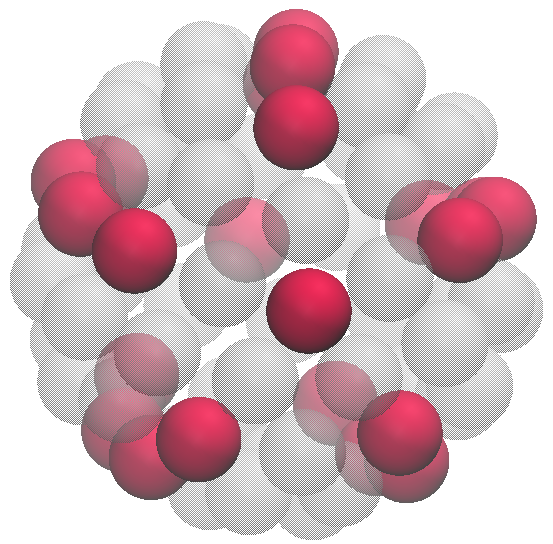}
\caption{PSD model, a couple of minimum-energy configurations:
$(1,1,1)$ interaction and $N=36$ (left); $(1,1,0)$ interaction and $N=22$ (right).
In this and similar pictures, particles (occupied sites) are represented as small red balls; 
holes (empty sites) as transparent light-gray balls.}
\label{111e110}
\end{figure}

Looking at Fig.~\ref{fig1}, we easily realize that two occupied fivefold sites are always left uncoupled by Hamiltonian (\ref{eq1}), opening to the possibility of stabilizing a regular icosahedral arrangement of zero energy when the interaction is an extended repulsion.
This is indeed observed, e.g., in the $(1,1,1)$ system at $\mu\simeq 0.30$.
At higher $\mu$, we see the formation of worm-like structures;
for example, the lowest-energy configuration for $N=36$ contains three equivalent chains of particles, each resembling a two-headed worm (see Fig.~\ref{111e110} left).
As $\mu$ is increased further, the grid gets eventually completely filled.

\begin{figure}   \includegraphics[width=5.8cm]{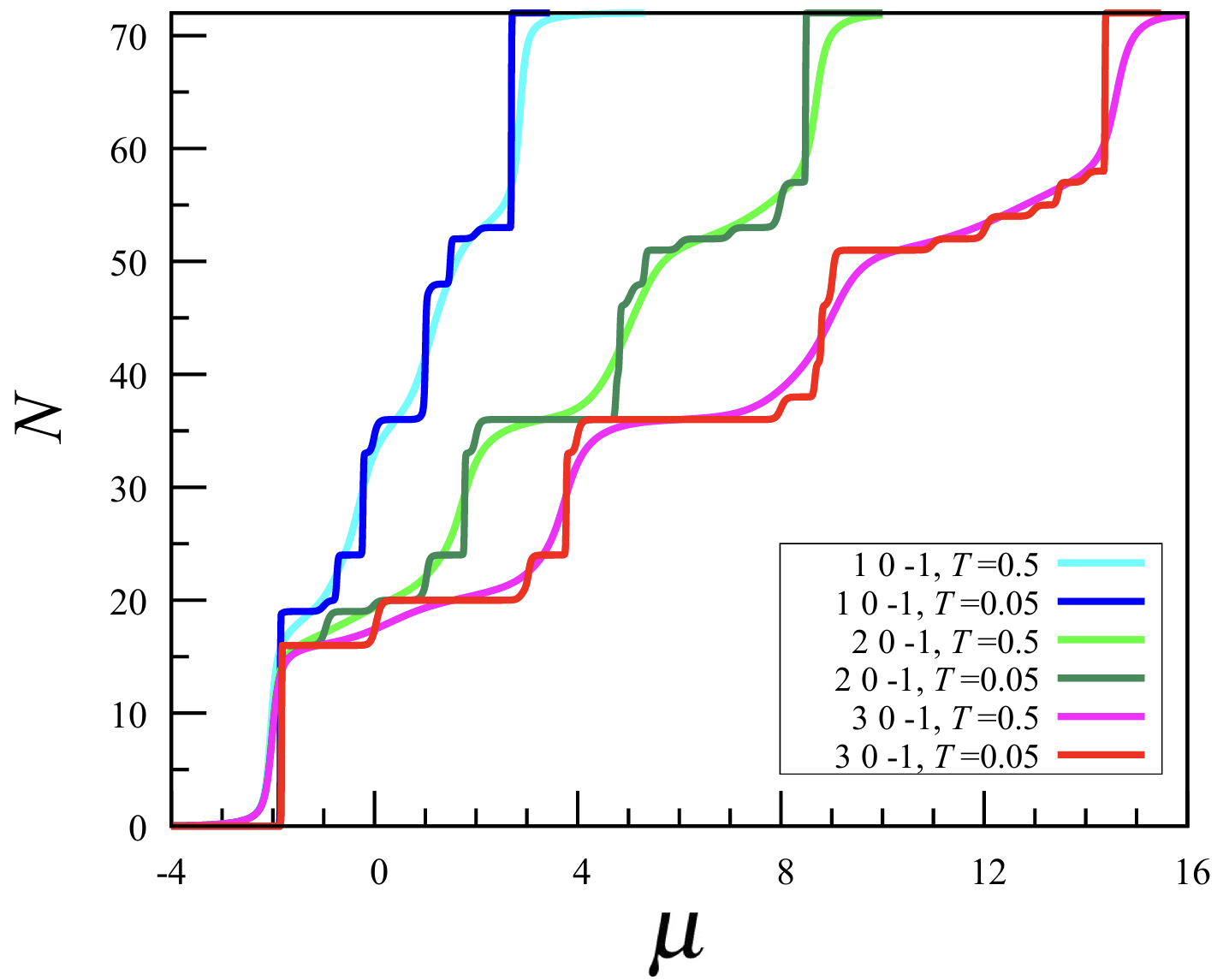}
\caption{PSD model, $(n,0,-1)$ interactions with $n=1,2,3$:
$N$ vs. $\mu$ for two temperatures (see legend).}
\label{psdx0-1}
\end{figure}

\begin{figure}
\centering
\includegraphics[width=3.5cm]{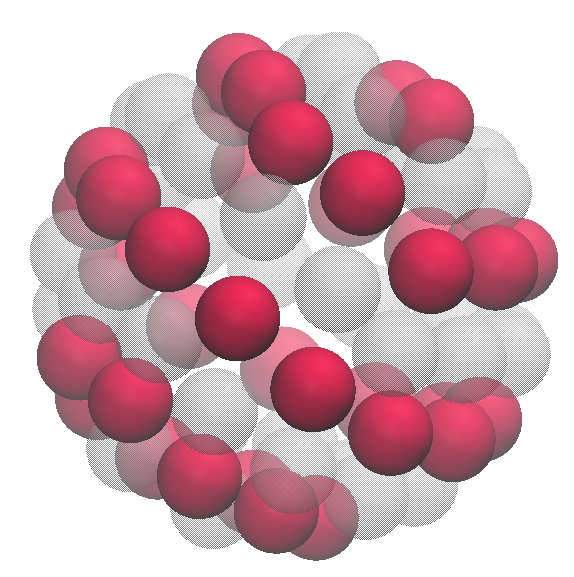}\,\,\,\,\,
\includegraphics[width=3.5cm]{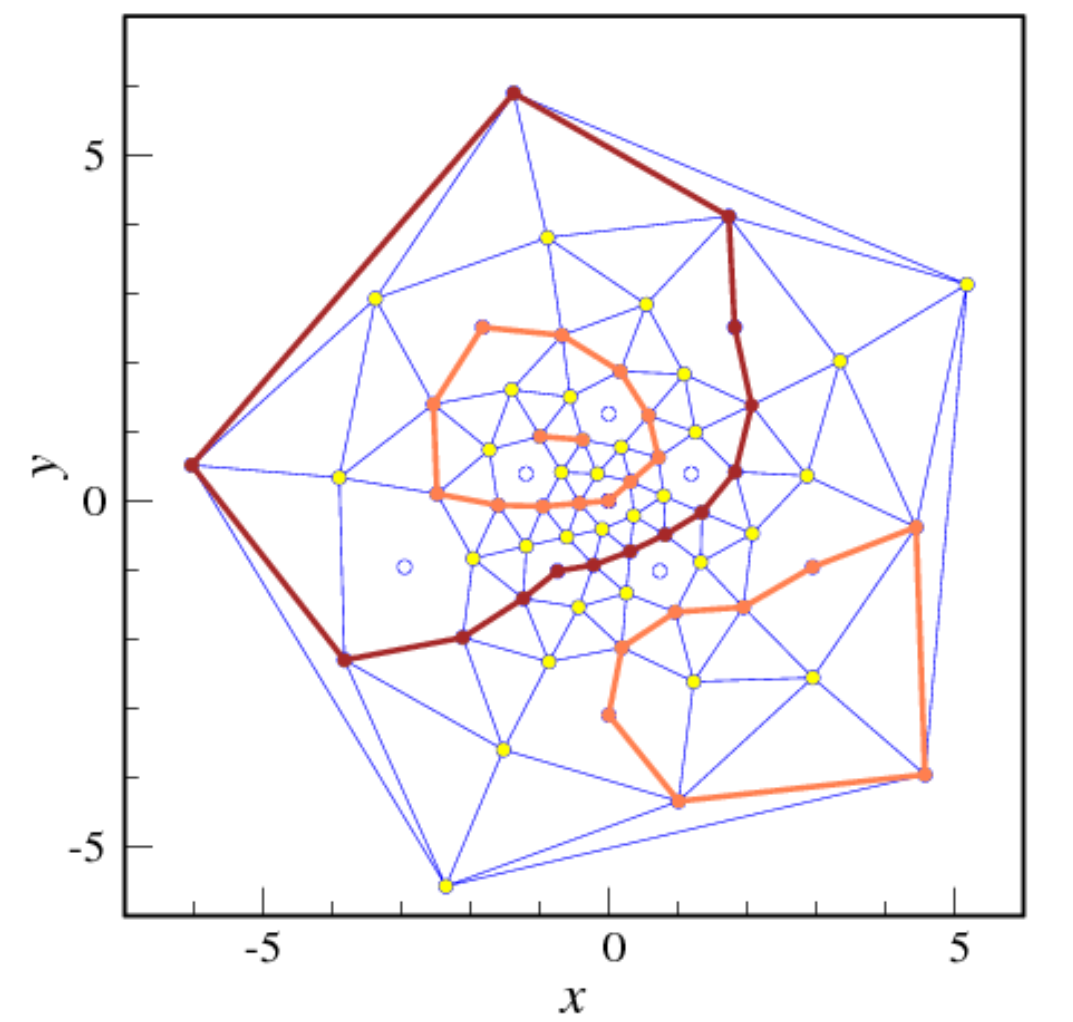}
\\
\includegraphics[width=3.5cm]{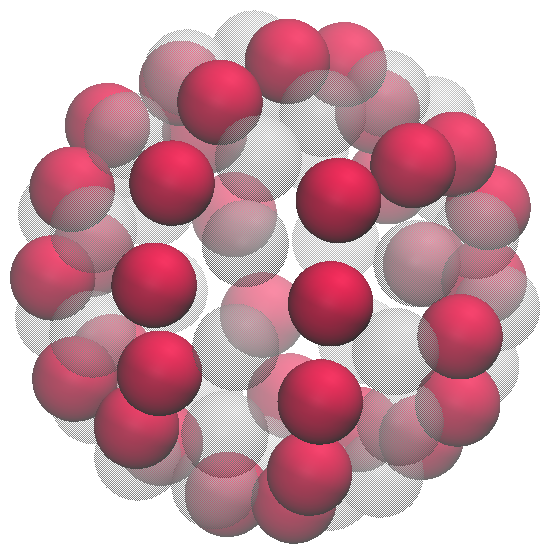}\,\,\,\,\,
\includegraphics[width=3.5cm]{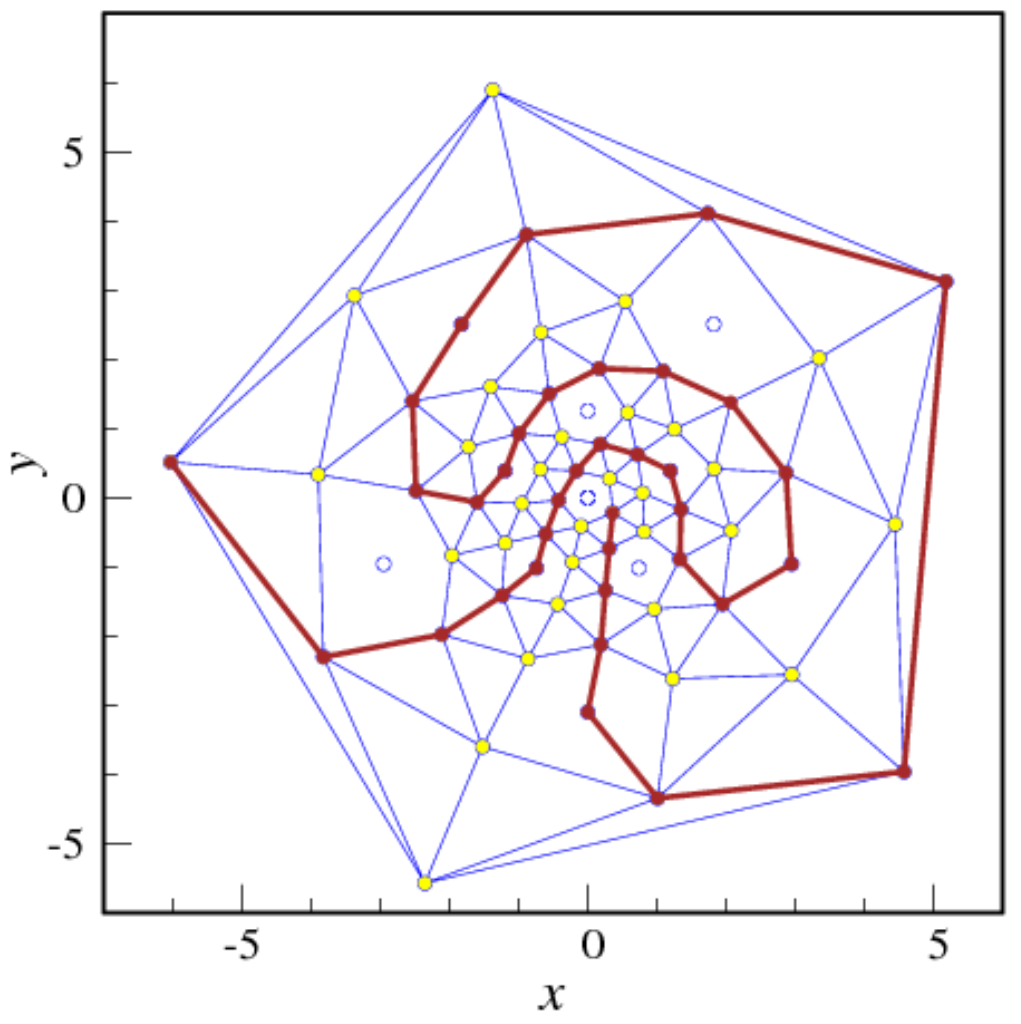}
\caption{PSD model, $(n,0,-1)$ interaction: The two minimum-energy configurations for $N=36$ (left), along with their stereographic projections (right).
The biggest cluster is colored in brown; smaller clusters are drawn in orange.
The horizontal and vertical scales have no particular meaning.}
\label{schlegel}
\end{figure}

\begin{figure}
\centering
\includegraphics[width=0.45\linewidth]{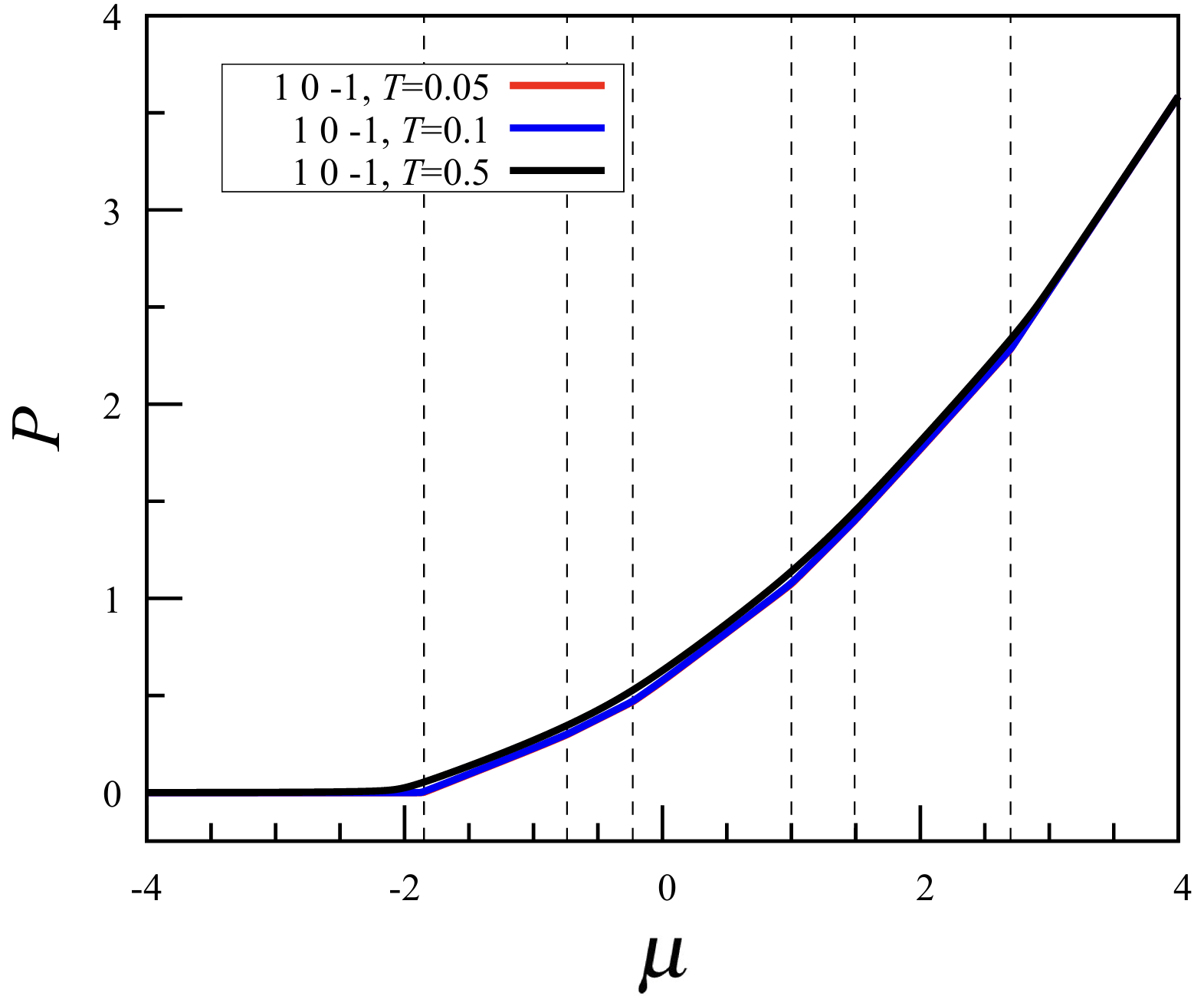}\,\,\,\,
\includegraphics[width=0.45\linewidth]{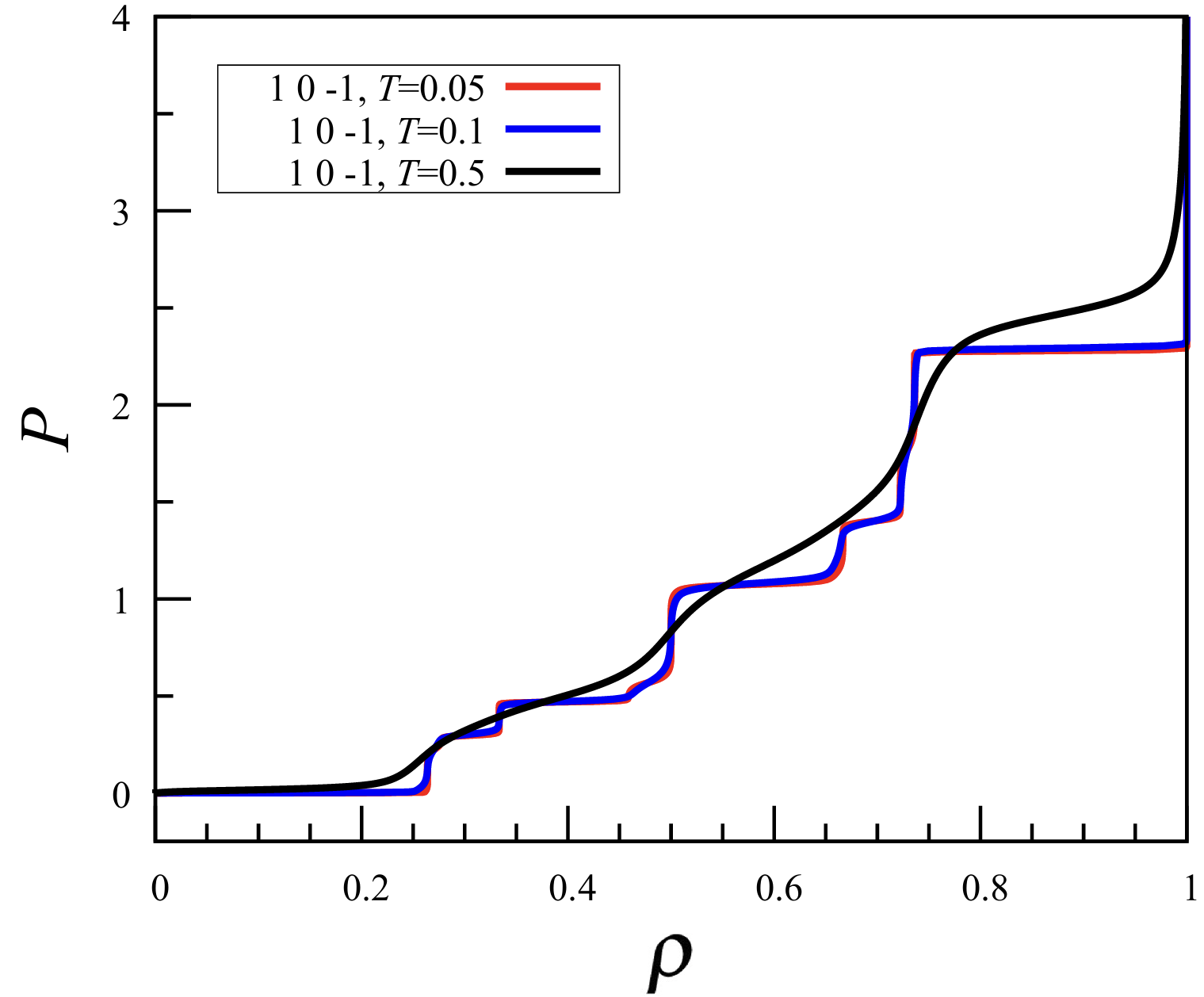}
\caption{PSD model, $(1,0,-1)$ interaction: $P$ vs. $\mu$ (left) and $P$ vs. $\rho$ (right) for three temperatures (see legend).}
\label{psd10-1P}
\end{figure}

\begin{figure}
\includegraphics[width=7cm]{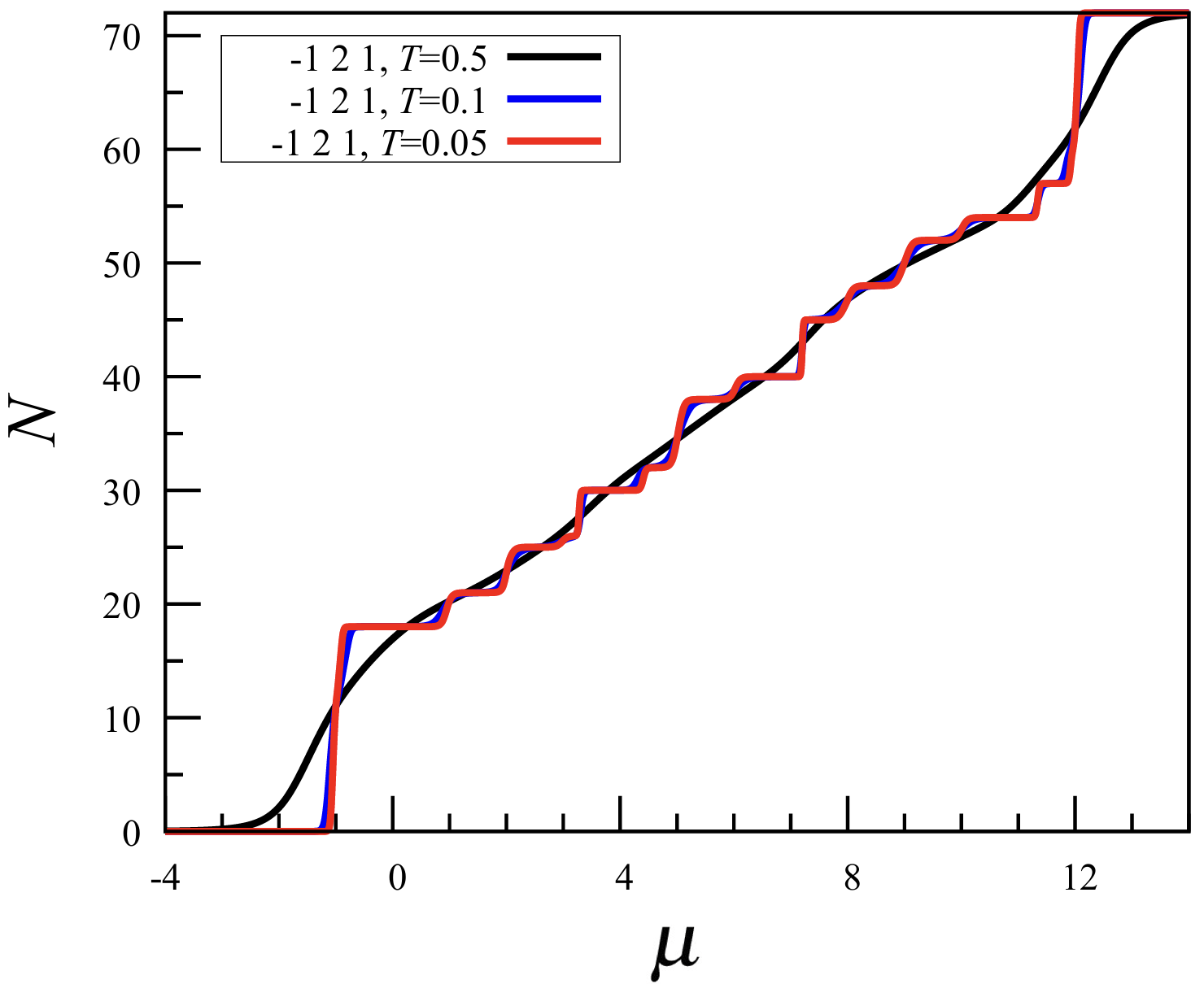}
\caption{PSD model, $(-1,2,1)$ interaction:
$N$ vs. $\mu$ for three temperatures (see legend).}
\label{psd-121N}
\end{figure}

\begin{figure}
\includegraphics[width=2.6cm]{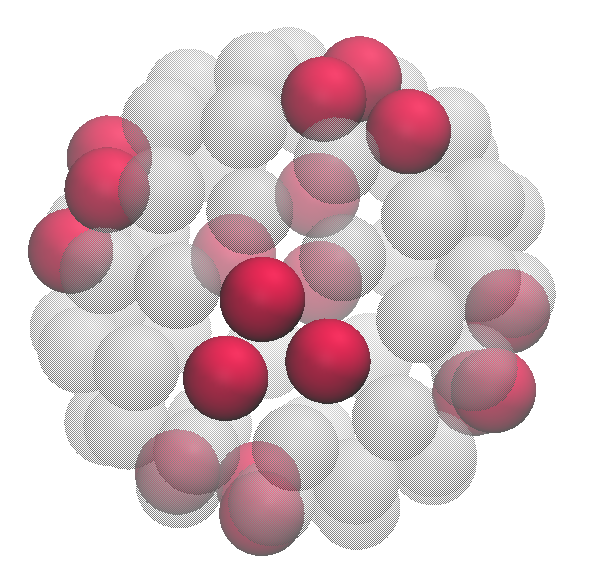}
\,\,
\includegraphics[width=2.6cm]{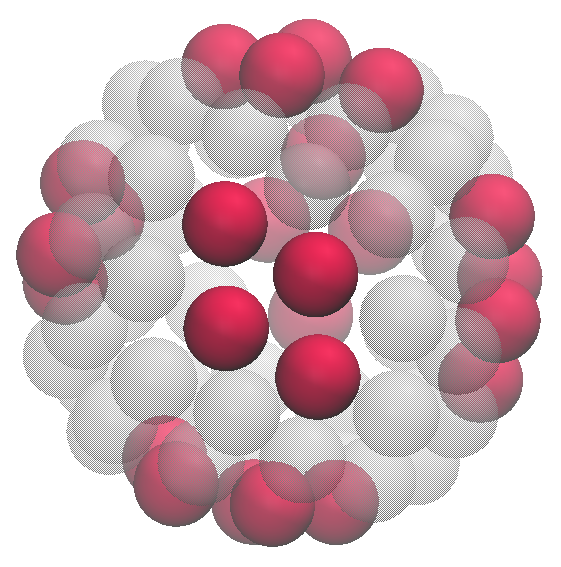}
\,\,
\includegraphics[width=2.6cm]{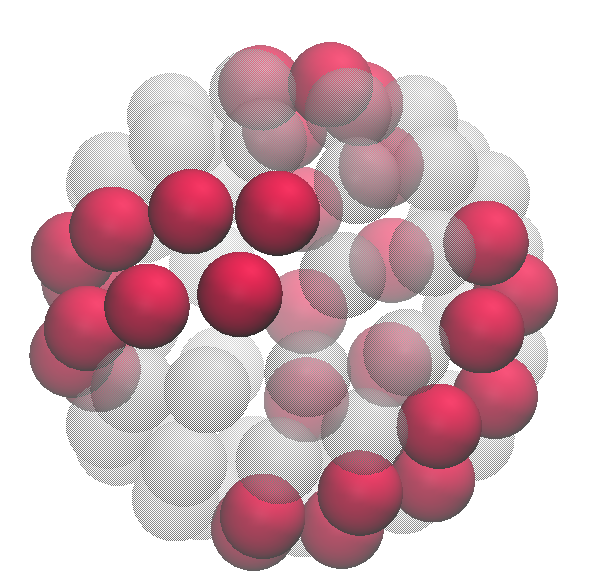}
\\
\includegraphics[width=2.6cm]{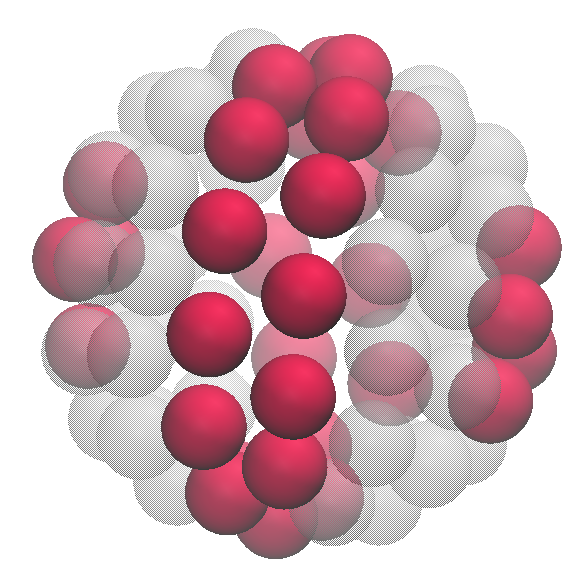}
\,\,
\includegraphics[width=2.6cm]{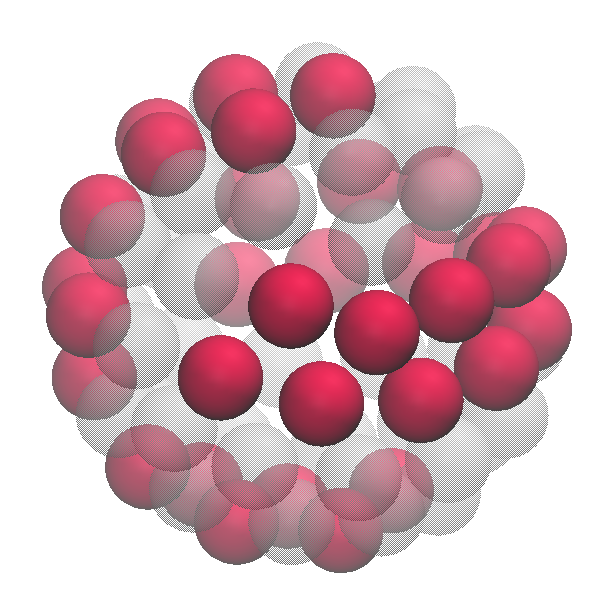}
\,\,
\includegraphics[width=2.6cm]{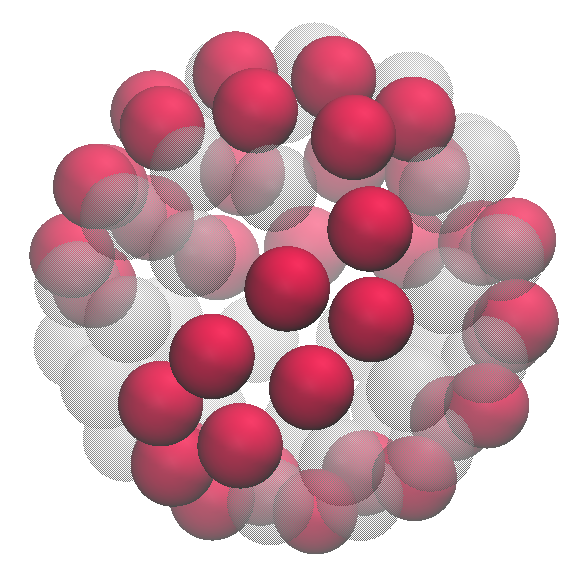}
\caption{PSD model, $(-1,2,1)$ interaction: from left to right and from top to bottom, minimum-energy configurations for  $N=18,25,30,32$ (two distinct), and 38.}
\label{psd-121}
\end{figure}

\begin{figure}
\includegraphics[width=2.6cm]{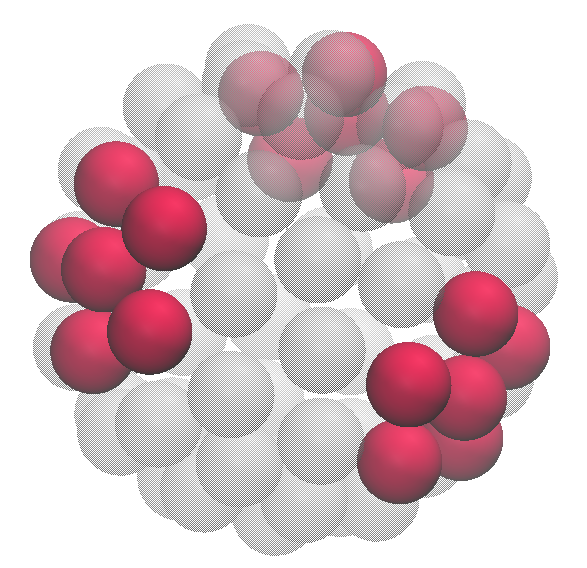}
\,\,
\includegraphics[width=2.6cm]{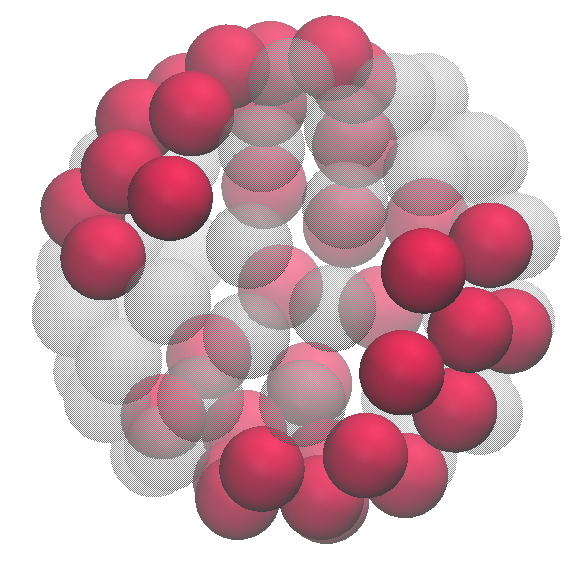}
\,\,
\includegraphics[width=2.6cm]{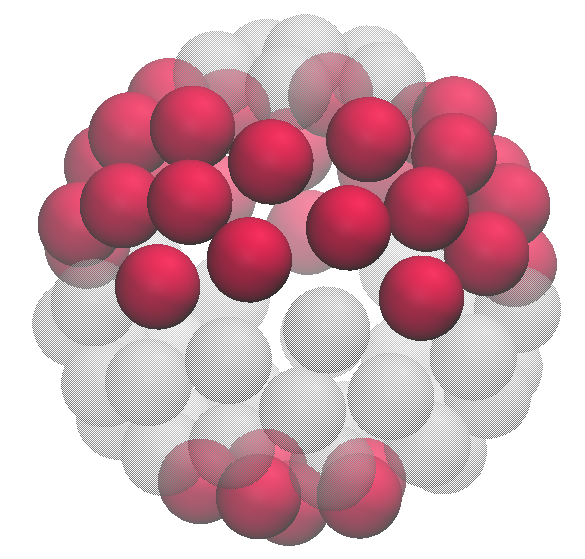}
\caption{PSD model, $(-1,0,1)$ interaction: minimum-energy configurations for $N=18$ and 36 (two distinct).}
\label{psd-101}
\end{figure}

Upon reducing the range of repulsion, other worm-like arrangements appear at low temperature (see an example in the right panel of Fig.~\ref{111e110}, which refers to the $(1,1,0)$ case).

When a third-neighbor attraction is added to a first-neighbor repulsion, as in a discretized Lennard-Jones potential, the equilibrium behavior changes completely.
In Fig.~\ref{psdx0-1} we show the profile of $N(\mu)$ for varying $u_1$ and fixed $u_2=0$ and $u_3=-1$.
At $T=0.05$ we see a sequence of plateaus separated by sharp crossovers, which become increasingly smoother as $T$ grows.
Each plateau represents a distinct ``phase''.
Strengthening the repulsion has just the effect of shifting the phase ``transitions'' to higher $\mu$.

More specifically, for $\mu<-2$ the grid is empty.
For higher chemical potentials, the system goes across a number of phases characterized by particle chains of growing length.
When $N=36$ the minimum-energy configurations are two:
one characterized by stripes and another showing a single ``worm'' wrapped around the sphere (see Fig.~\ref{schlegel}).
Admittedly, the mechanism stabilizing both structures is a high number of third-neighbor particle pairs, which, in spite of the large number of adjacent particles in the chains, makes the total energy lower than in any other arrangement of 36 particles on the grid.
Figure 4 is a significant example of how a simple interaction can trigger the onset, out of an abundance of options, of mesoscopic structures with an easily recognizable shape.
Other examples follow below.

Looking at the profile of the isothermal compressibility (not shown), a sequence of peaks appears at low temperature, aligned with the jumps present in $N(\mu)$.
For the $(1,0,-1)$ interaction the pressure $P(\mu)$ is reported in Fig.~\ref{psd10-1P} left, where we have marked with dashed lines the location of the $\kappa_T$ peaks. 
The right panel of Fig.~\ref{psd10-1P} shows the equation of state, $P(\rho)$, for the same three temperatures.
Characteristic plateaus are clearly visible in $P(\rho)$ at low temperature, as would be expected for a system undergoing a sequence of first-order transitions.

The $(n,0,-1)$ interaction is not the only one to exhibit stripes and other worm-like structures.
Indeed, all interactions characterized by a first-neighbor repulsion and a minimum falling at third-neighbor distance display a behavior similar to $(1,0,-1)$, with slight variations in the intermediate phases.

Things change considerably when combining a first-neighbor repulsion with a second-neighbor attraction and a weaker (or zero) third-neighbor attraction.
In this case, more irregular patterns emerge, causing stripes and worms to disappear.
As previously discussed, another way of realizing a Lennard-Jones-like interaction is to take a negative $u_1$ and less negative (or zero) $u_2$ and $u_3$.
For instance, in the $(-1,0,0)$ case the profile of $N(\mu)$ jumps abruptly from $\approx 0$ to 72, as it would happen in a system undergoing a transition from vapor to solid.
This is confirmed by the shape of the equation of state having a unique plateau (not shown).

Upon providing a first-neighbor attraction with a repulsive tail we realize the lattice-gas equivalent of a SALR potential.
The short-range attraction promotes particle aggregation, while the longer-range repulsion prevents the immediate formation, with increasing $\mu$, of a solid-like phase.
A non-trivial behavior emerges for, e.g., $u_1=-1$, $u_2=2$, and $u_3=1$, where the interplay between attraction and repulsion leads the system to explore a whole sequence of cluster-crystal microphases of increasing complexity (see Figs.~\ref{psd-121N} and \ref{psd-121}).
In the first two snapshots of Fig.~\ref{psd-121} (corresponding to $N=18$ and $N=25$) the clusters are centered at the vertices of an octahedron.
Notice, in particular, how each cluster is built around a fivefold vertex.
For $N=30$ we see the emergence of three elongated clusters, consisting of $10$ particles each.
For $N=32$ the minimum-energy configurations are two:
one featuring a ring of particles along the equator, accompanied by four-particle clusters at poles, and another exhibiting a characteristic ``tennis ball'' pattern.
Finally, for $N=38$ we observe a long curved ribbon ``wandering'' through the grid.
As $N$ increases further, we find a number of low-energy configurations that are complementary to the ones just described, even though no perfect ``particle-hole'' symmetry holds in this case.
All in all, the above scenario can be viewed as the lattice-gas version of the generic phase diagram of a two-dimensional SALR fluid at low temperature (see, e.g., Refs.~\cite{Ciach2018,Pini2015}).
Structures very similar to those illustrated in Fig.~\ref{psd-121} are found in the $(-1,2,0)$ system too.

Another intriguing SALR interaction is $(-1,0,1)$.
While the overall low-temperature behavior is similar to $(-1,2,1)$, the relevant structures are different.
As shown in Fig.~\ref{psd-101}, in the $N=18$ phase we find three pentagon-shaped clusters of six particles each, centered on fivefold vertices.
For $N=36$ two different lowest-energy configurations exist: 
one formed by a curved ribbon and another by two unequal clusters.

\subsection{HPTI model}

Using the same order of presentation as before, we begin our analysis of the HPTI model by core-corona repulsions.
Again, our interest is mainly in the most stable structures, which are those heralded by a fairly extended plateau in $N(\mu)$ at low $T$.

Taking $u_1=1$ and $u_2=u_3=0$, the most distinct phase appears at $N=32$, where particles are arranged on the vertices of a pentakis dodecahedron; in this case, the empty sites form its dual polyhedron, i.e., the truncated icosahedron. 
Extending the repulsion range a little further, the system goes through a sequence of mostly disordered phases, where small clusters coexist with short particle chains.
Finally moving to $(1,1,1)$, a broader variety of structures emerges.
For low $\mu$ values, $N=12$ particles are organized into an icosahedral configuration, either regular or irregular.
As the density increases, we see small clusters mixed up with worm-like structures.

\begin{figure}
\centering
\includegraphics[width=0.45\linewidth]{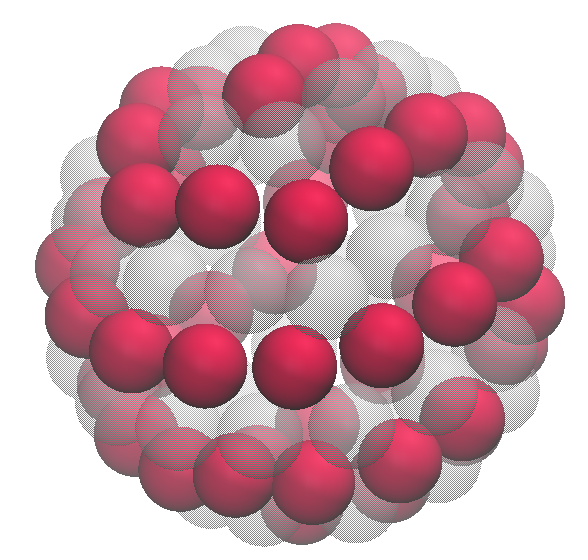}\,\,\,\,\,
\includegraphics[width=0.45\linewidth]{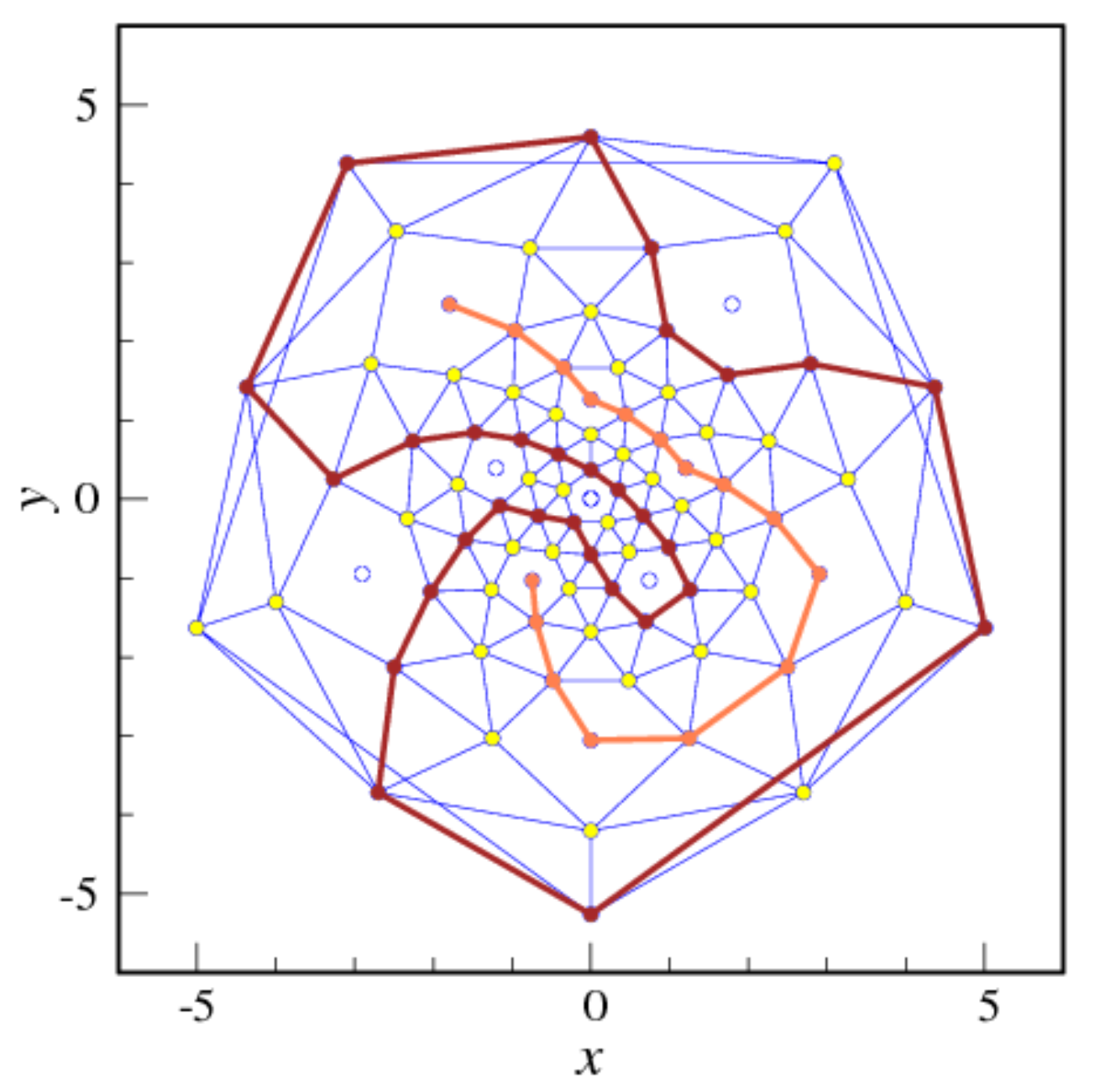}
\caption{HPTI model, $(2,0,-1)$ interaction:
minimum-energy configuration for $N=46$ (left), along its stereographic projection (right).}
\label{hpti20-1N}
\end{figure}

\begin{figure}
\centering
\includegraphics[width=0.45\linewidth]{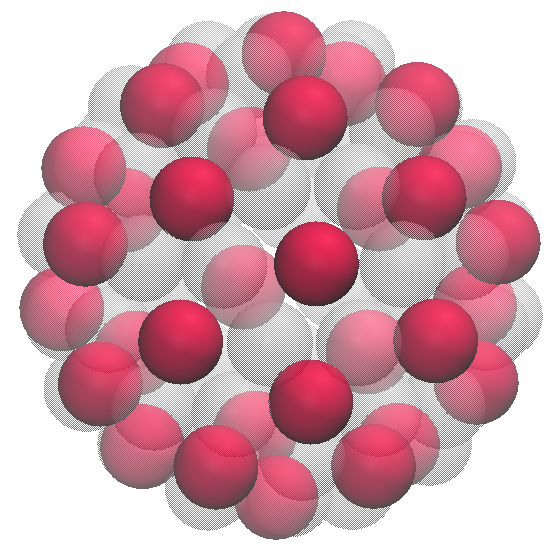}\,\,\,\,\,
\includegraphics[width=0.45\linewidth]{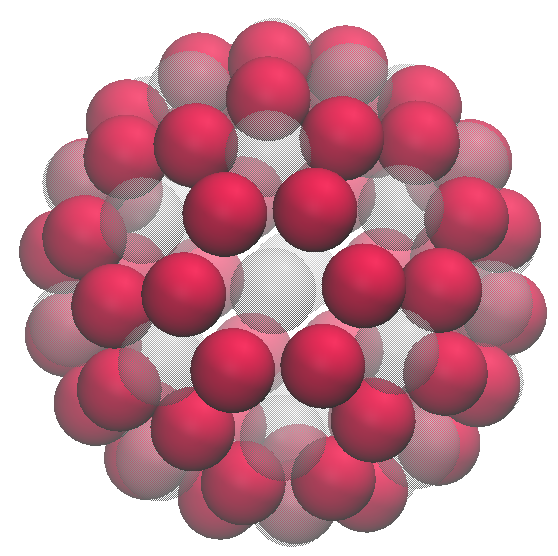}
\caption{HPTI model, $(1,-1,0)$ interaction:
pentakis dodecahedron ($N=32$, left) and truncated icosahedron ($N=60$, right).}
\label{hpti1-1}
\end{figure}

Similarly as done for the PSD model, we examine three variants of Lennard-Jones-like interaction.
Considering first the $(2,0,-1)$ case, the third-neighbor attraction leads to the formation of a disordered arrangement of isolated particles at low density ($N=22$).
As the density increases ($N=30$), small worm-like structures emerge.
For $N=46$ (see Fig.~\ref{hpti20-1N}) the minimum-energy configuration contains two worms wrapped around each other, which subsequently merge into a single percolating worm at $N=48$.
At still higher densities, we observe the structures complementary to those found at $N=22$ and $N=30$.

\begin{figure}
\includegraphics[width=3.7cm]{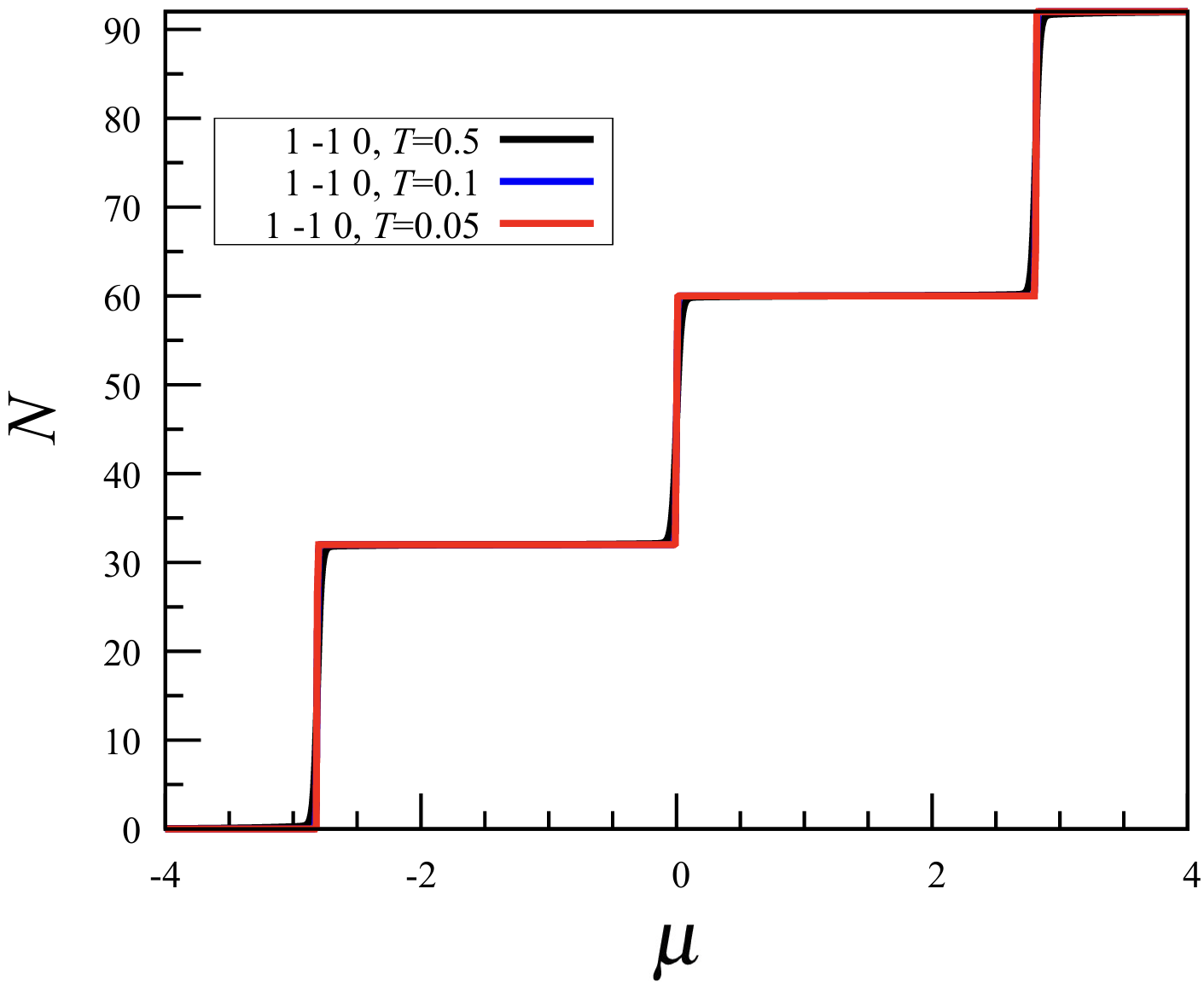}\,\,\,\,\,
\includegraphics[width=3.7cm]{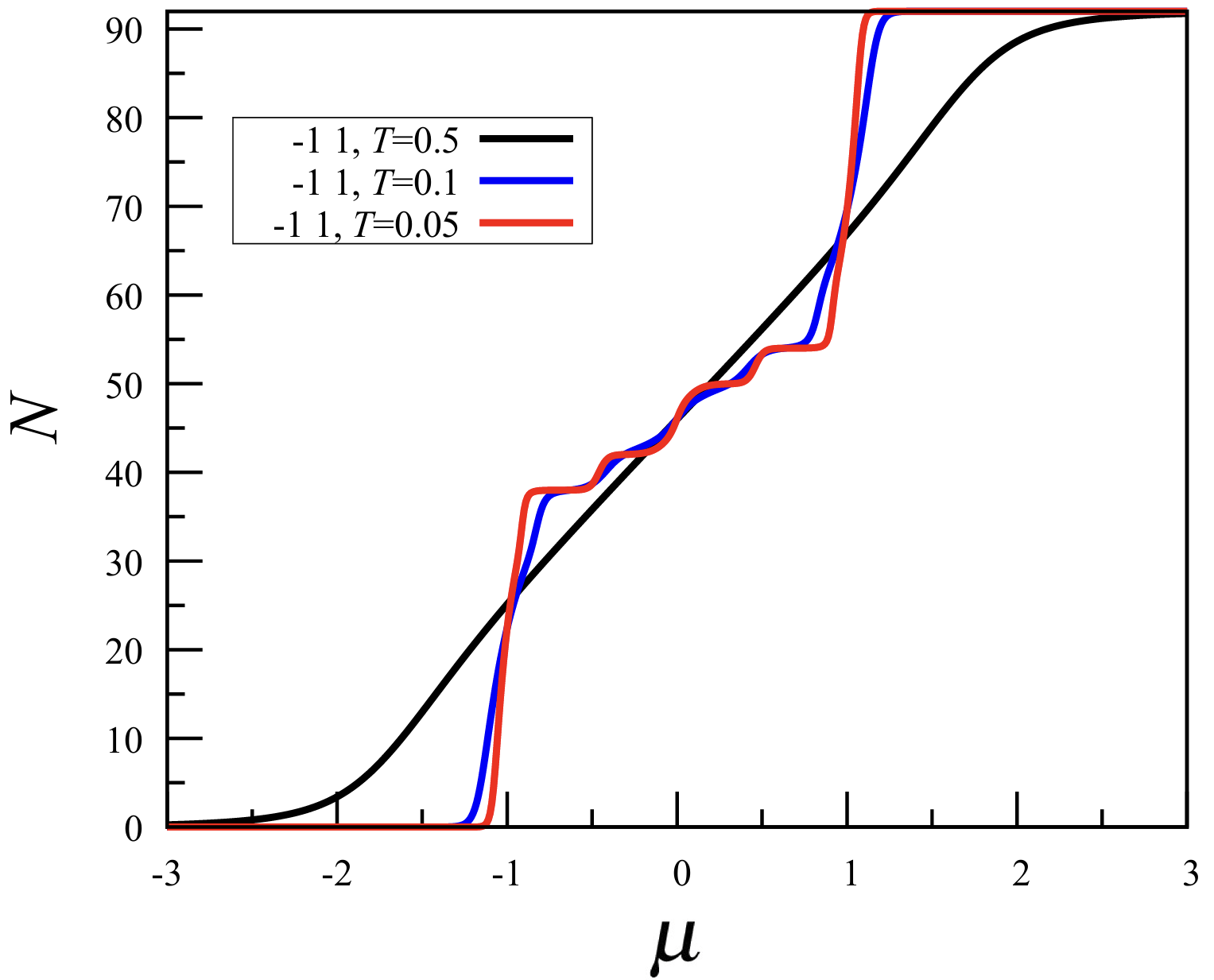}
\caption{HPTI model, $N(\mu)$ for two  sets of parameters with $u_1+u_2=0$ and $u_3=0$ and three temperatures (in the legend):
$(1,-1,0)$ (left) and $(-1,1,0)$ (right).
Particle-hole symmetry holds for both models (see Appendix A).}
\label{symmetries}
\end{figure}

When the first-neighbor repulsion is weakened, e.g., $(1,0,-1)$, the previous $N=46$ phase disappears.
Conversely, increasing the repulsion leads to the emergence of additional intermediate phases, which however are poorly significant.
If we put the minimum of the interaction at second-neighbor distance, with a null or weaker third neighbor attraction, the behavior of the system changes radically.
Specifically, the $(1,-1,0)$ system undergoes a sort of solid-solid transition between a $N=32$ phase, where particles form a pentakis dodecahedron, and a $N=60$ phase, where particles are placed at the vertices of a truncated icosahedron (see Fig.~\ref{hpti1-1}).
The $N(\mu)$ curve, reported in the left panel of Fig.~\ref{symmetries} for three temperatures, makes it evident that these two phases are extremely robust to thermal fluctuations, much more than any other ``phase'' encountered so far.
Furthermore, the profile of $N(\mu)$ is symmetric with respect to $\mu_0=0$ (this is not a mere coincidence; see Appendix A for a proof).

\begin{figure}
\includegraphics[width=2.6cm]{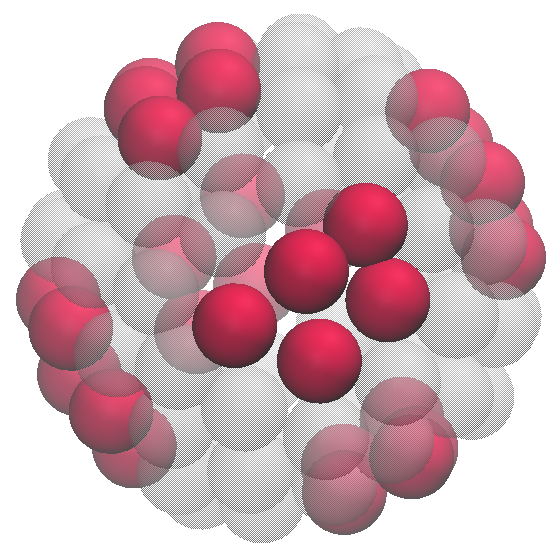}\,\,
\includegraphics[width=2.6cm]{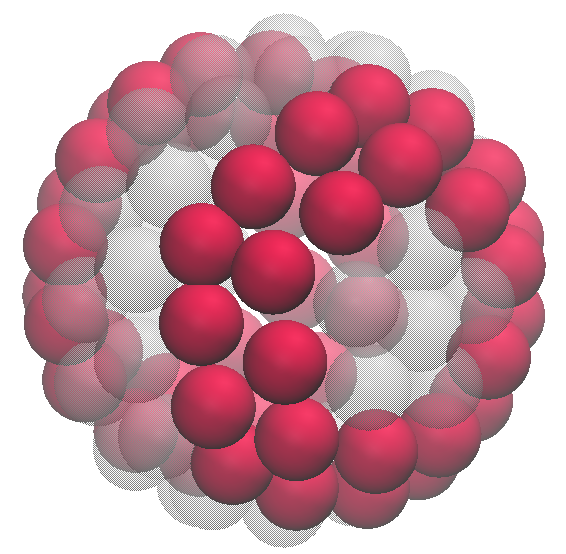}\,\,
\includegraphics[width=2.6cm]{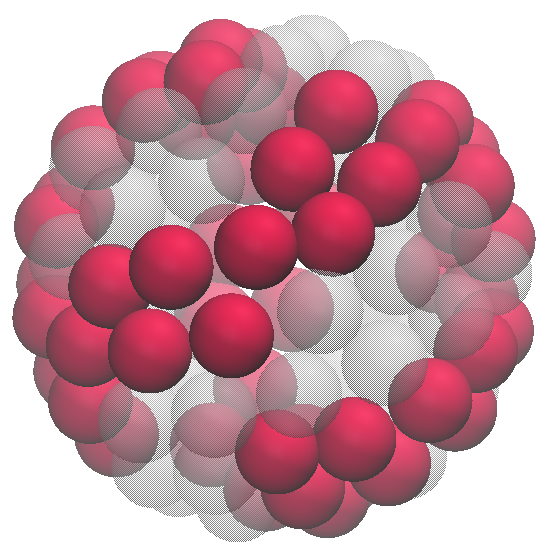}
\caption{HPTI model, ($-1,2,1$) interaction:
minimum-energy configurations for $N=30,50,52$.}
\label{hpti-121}
\end{figure}

Finally moving to SALR interactions, the parameter set $(-1,2,1)$ offers a wide variety of structures, as highlighted by the numerous small steps in $N(\mu)$ (not shown).
In particular, at low density the system undergoes a sequence of transitions between cluster-crystal configurations with octahedral symmetry around fivefold vertices (see Fig.~\ref{hpti-121}).
At higher densities clusters merge into thick worms, which first evolve to a ``tennis ball'' ($N=42$) and then to stripes ($N=46-50$). For $N=52$ a single worm finally appears.
A further increase in the density leads to configurations that are complementary to cluster crystals.

Reducing the height and range of the repulsion ramp, e.g., $(-1,1,0)$, the system exhibits two distinct phases, along with their complementary phases (see $N(\mu)$ in the right panel of Fig.~\ref{symmetries}).
Both configurations show a ring of particles along the equator, with either four ($N=38$) or six ($N=42$) particles at opposite poles.

\subsection{HPCD model}

\begin{figure}
\centering
\includegraphics[width=0.4\linewidth]{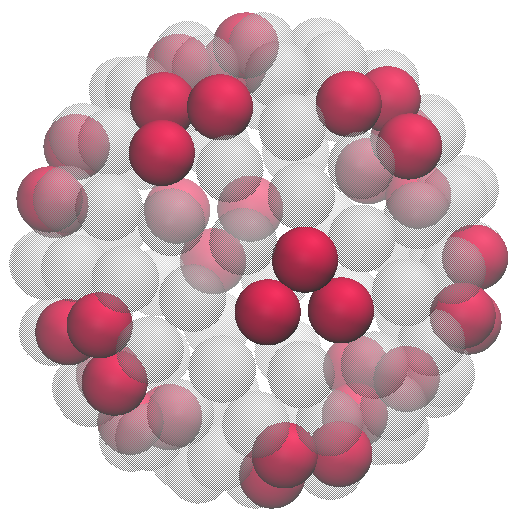}\,\,\,\,\,
\includegraphics[width=0.4\linewidth]{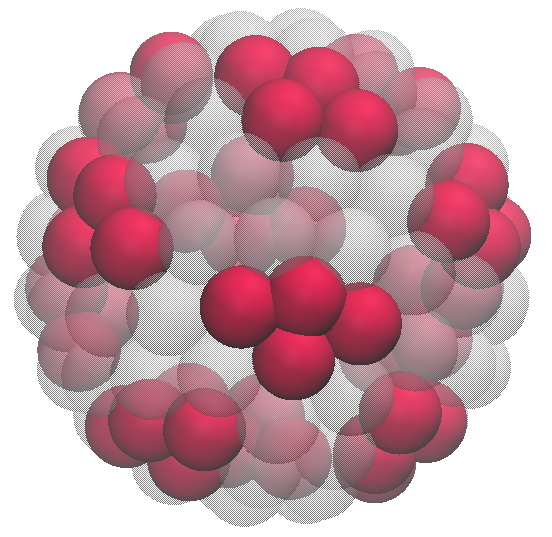}\\
\includegraphics[width=0.4\linewidth]{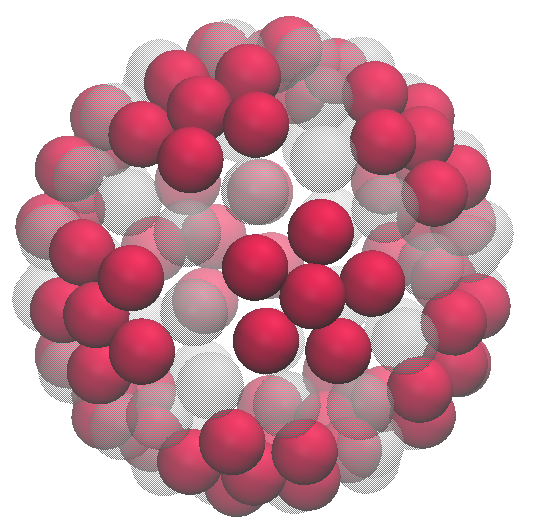}\,\,\,\,\,
\includegraphics[width=0.4\linewidth]{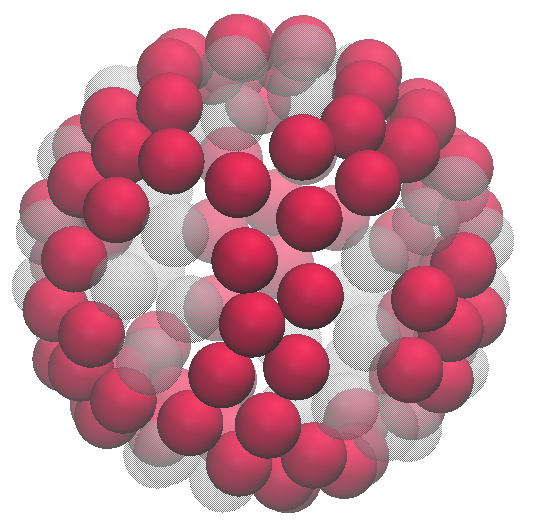}
\caption{HPCD model, $(1,1,1)$ interaction: minimum-energy configurations for $N=36,48,72,82$ (from left to right, from top to bottom).}
\label{hpcd111}
\end{figure}

Again, our analysis of the HPCD model begins with core-corona interactions.
At low $T$, the $(1,0,0)$ system organizes itself in polyhedral structures.
Specifically, $N=42$ particles occupy the vertices of a pentakis icosidodecahedron, while two irregular polyhedra emerge at $N=78$ and $N=86$.
When the repulsion is extended to include second neighbors, the system behavior changes substantially.
Regular structures persist at low density, as highlighted by the presence of a $N=32$ pentakis dodecahedron phase, but, for intermediate densities, worm-like structures of various length appear, progressively merging into a connected branched arrangement of $N=64$ particles.
Moving to the $(1,1,1)$ interaction, we find a few cluster-crystal structures (see Fig.~\ref{hpcd111}).
In particular, we observe icosahedral arrangements of clusters comprised of three ($N=36$) or four particles ($N=48$), built around fivefold sites.
As already found for other grids, worms arise at intermediate density and coexist with clusters, until at $N=72$ we see the onset of twelve pentagon-shaped clusters of six particles each, arranged at the vertices of an icosahedron.
At $N=82$ we observe a complex yet symmetric arrangement of clusters and stripes, while at higher density we find a percolated assembly of particles with ``holes'' at the vertices of an icosahedron.
Finally, in the minimum-energy configuration with $N=108$ particles the empty sites occupy the vertices of a rhombic dodecahedron~\cite{visualpolyhedra}.

\begin{figure*}
\centering
\includegraphics[width=0.3\linewidth]{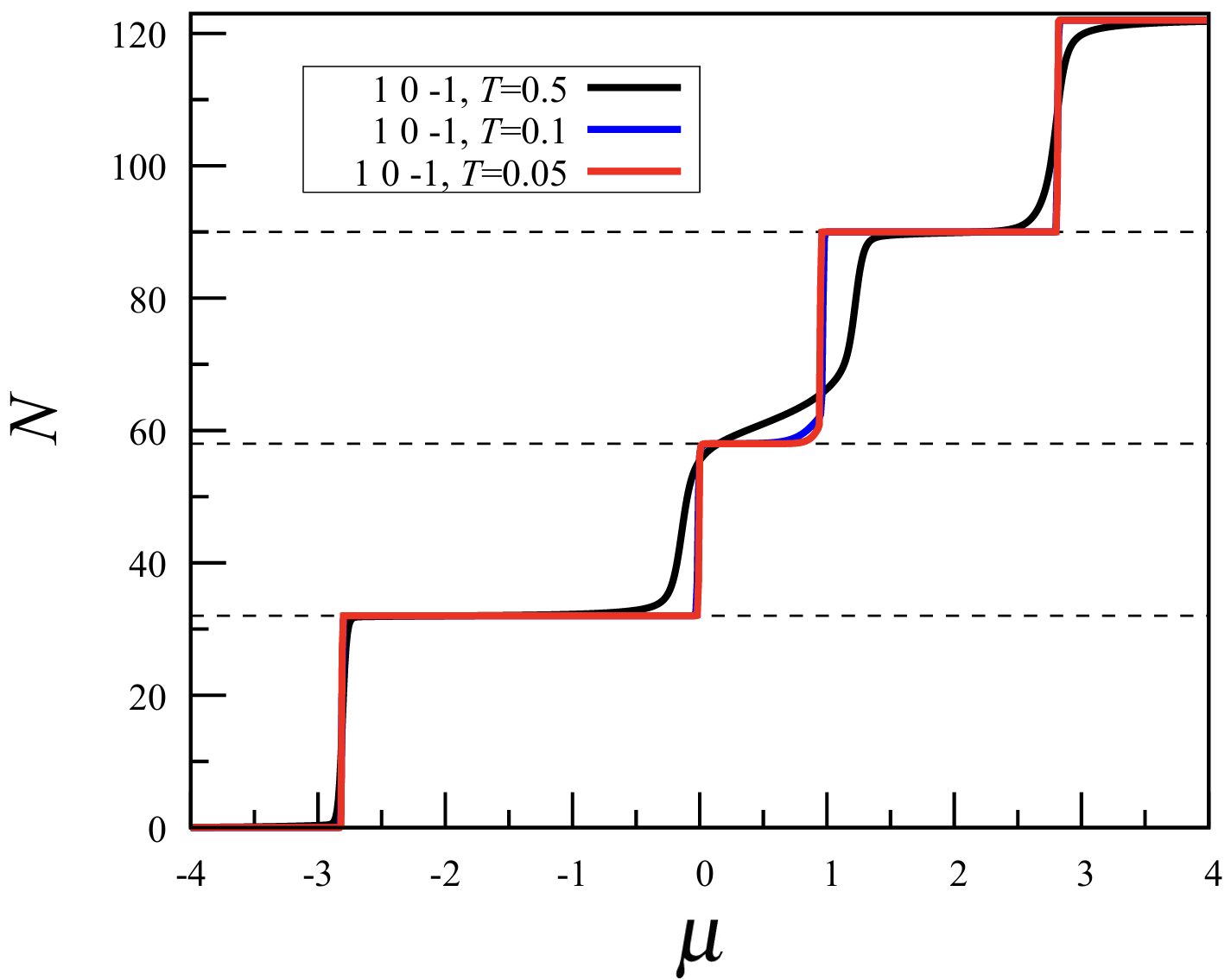}
\includegraphics[width=0.3\linewidth]{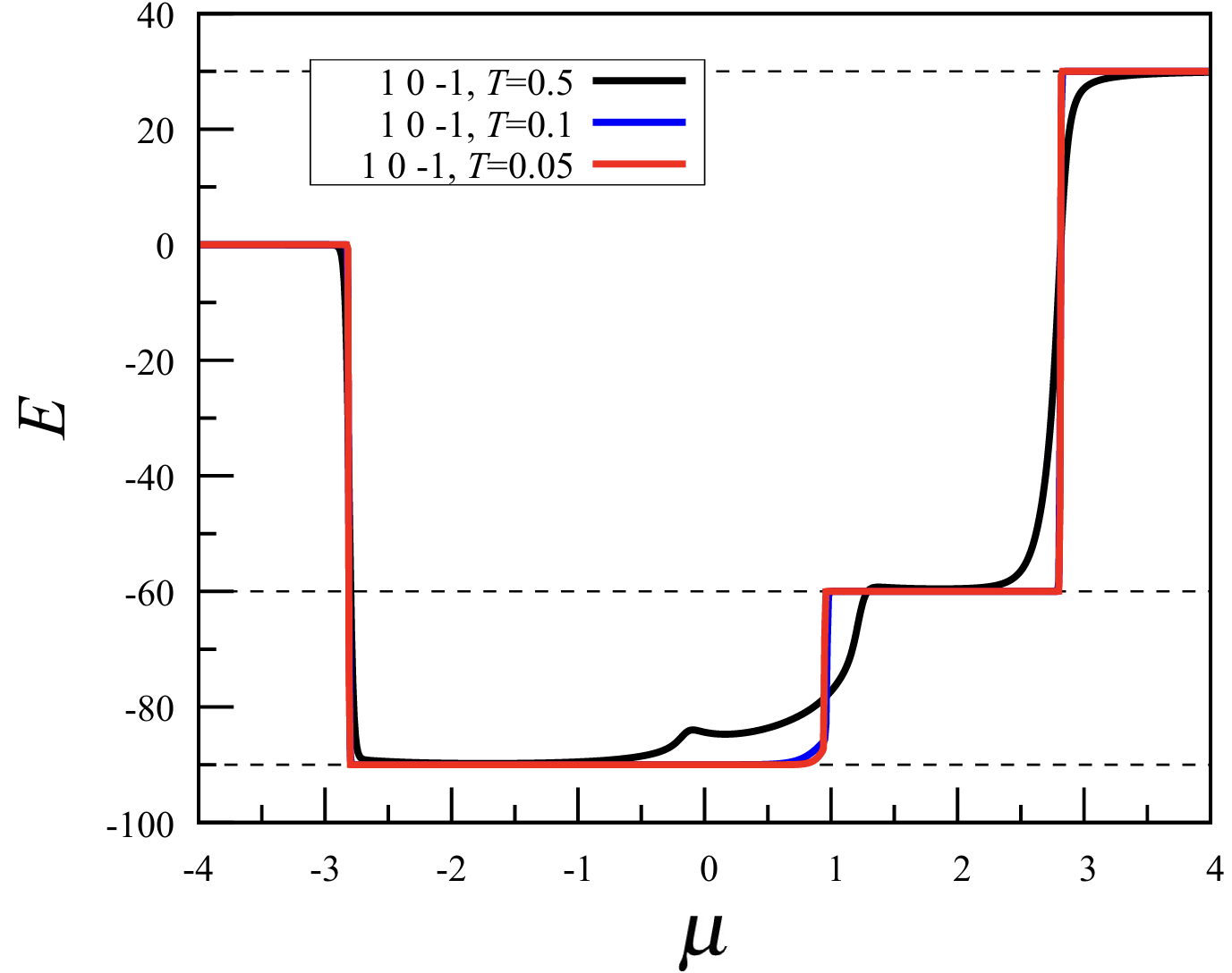}
\includegraphics[width=0.3\linewidth]{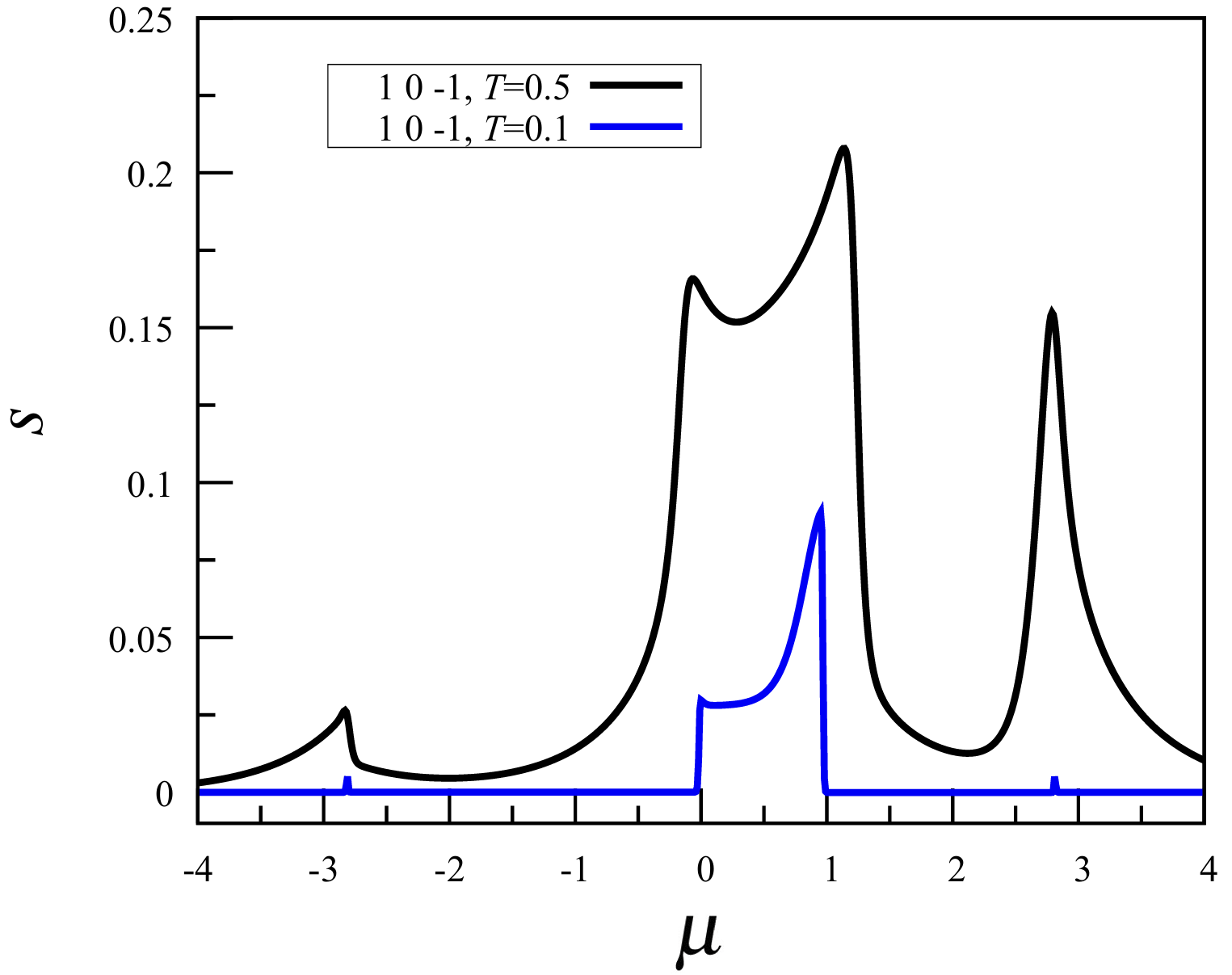}
\caption{HPCD model, $(1,0,-1)$ interaction: $N(\mu)$, $E(\mu)$, and $s(\mu)$.}
\label{hpcd10-1NES}
\end{figure*}

\begin{figure*}
\includegraphics[width=5.8cm]{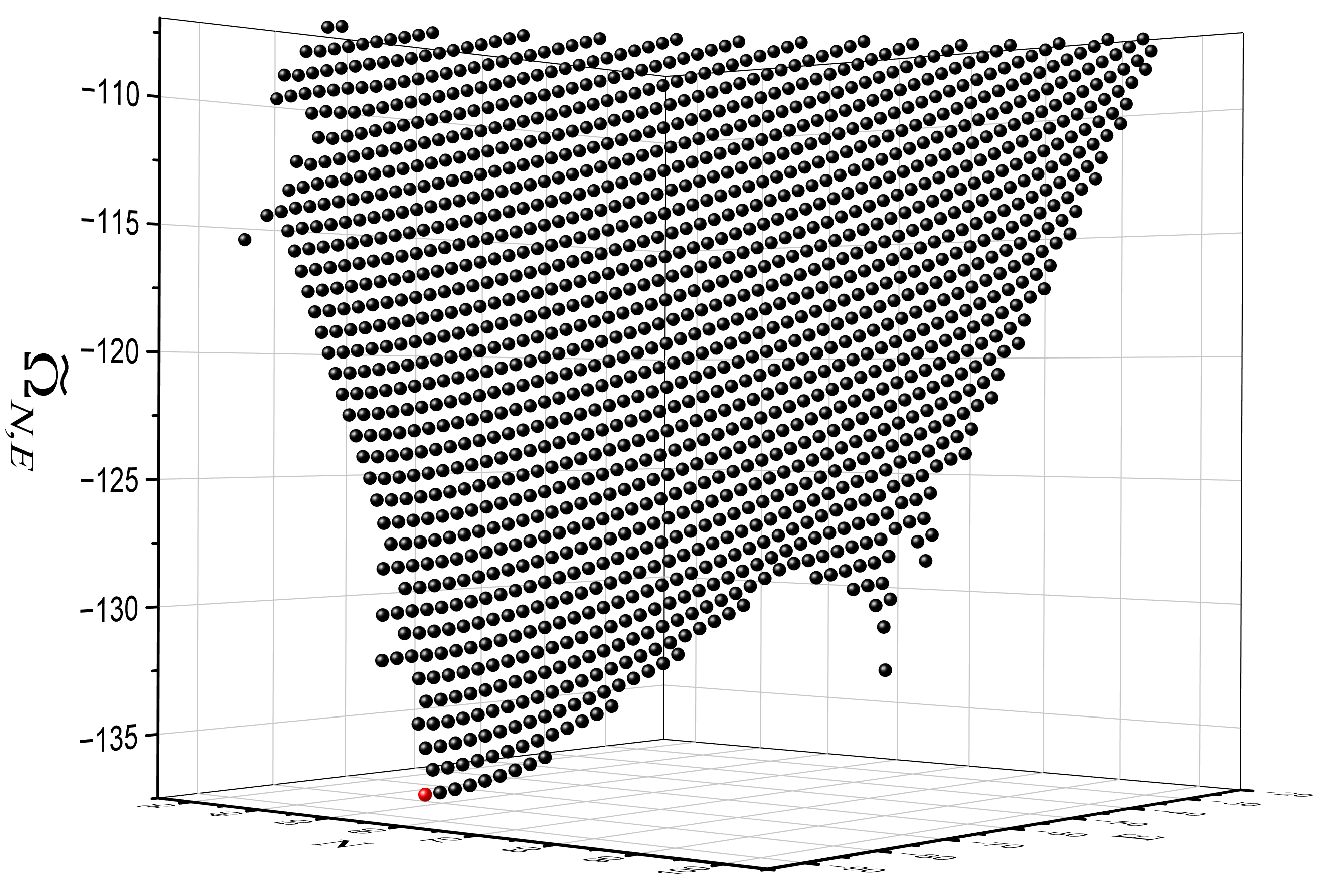}
\hfill
\includegraphics[width=5.8cm]{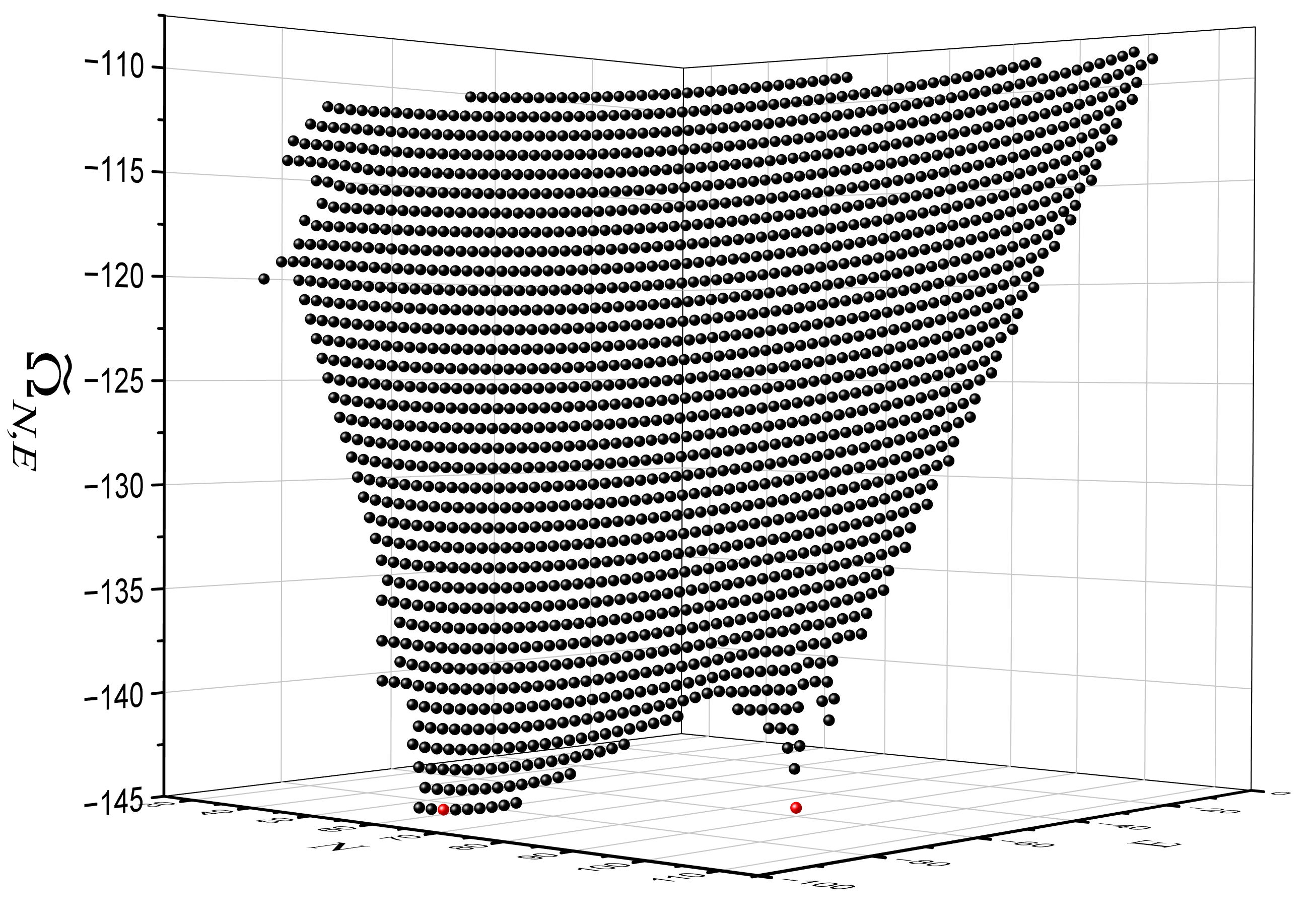}
\hfill
\includegraphics[width=5.8cm]{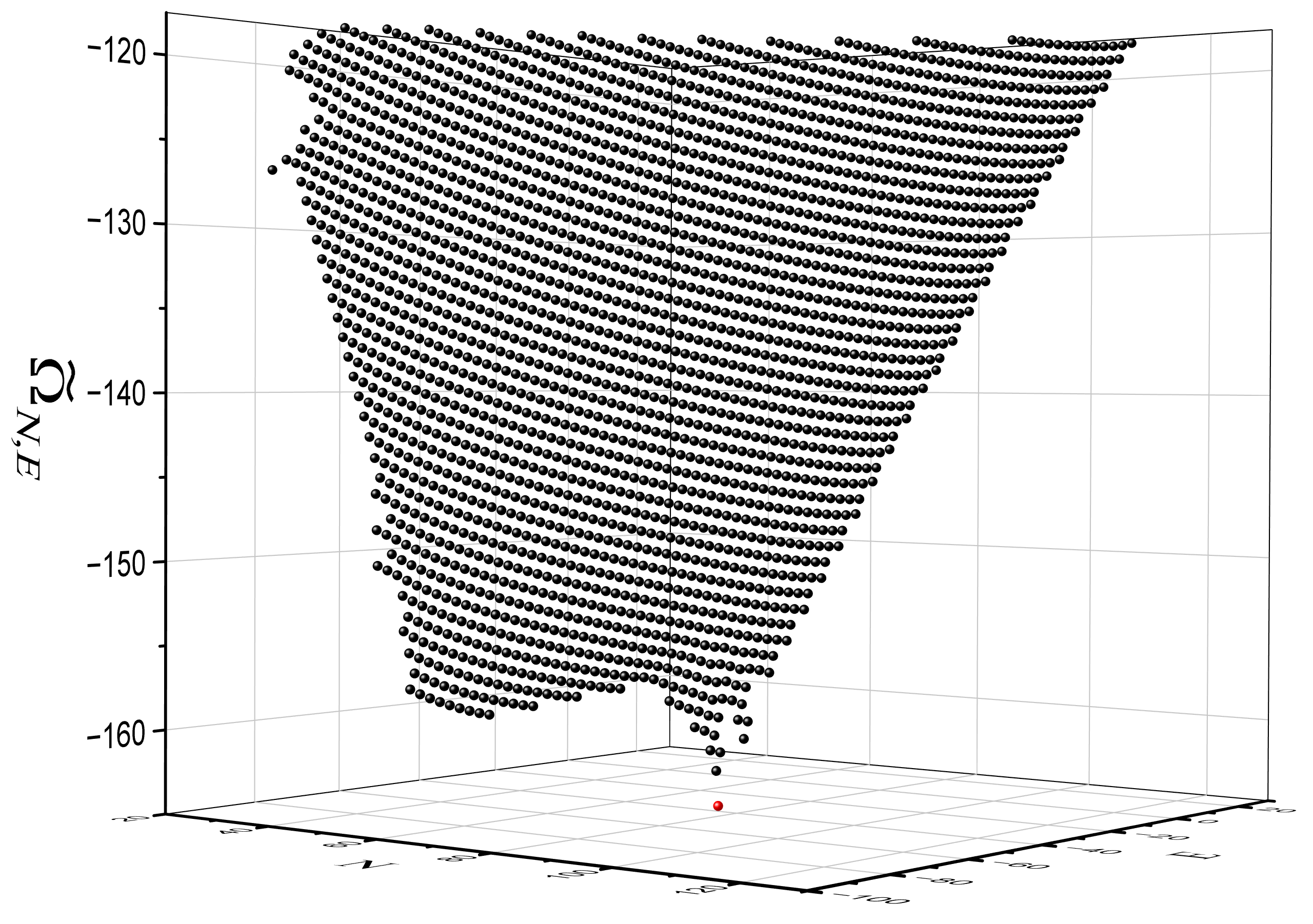}\,\,
\caption{HPCD model, $(1,0,-1)$ interaction:
GGP plotted as a function of ${\cal N}$ and ${\cal E}$ for three different chemical potentials, $\mu=0.8$ (left), $\mu=0.95$ (middle), and $\mu=1.15$ (right).
We have marked in red the absolute minimum of the GGP in each case.} \label{omegatilde}
\end{figure*}

\begin{figure}
\includegraphics[width=3.5cm]{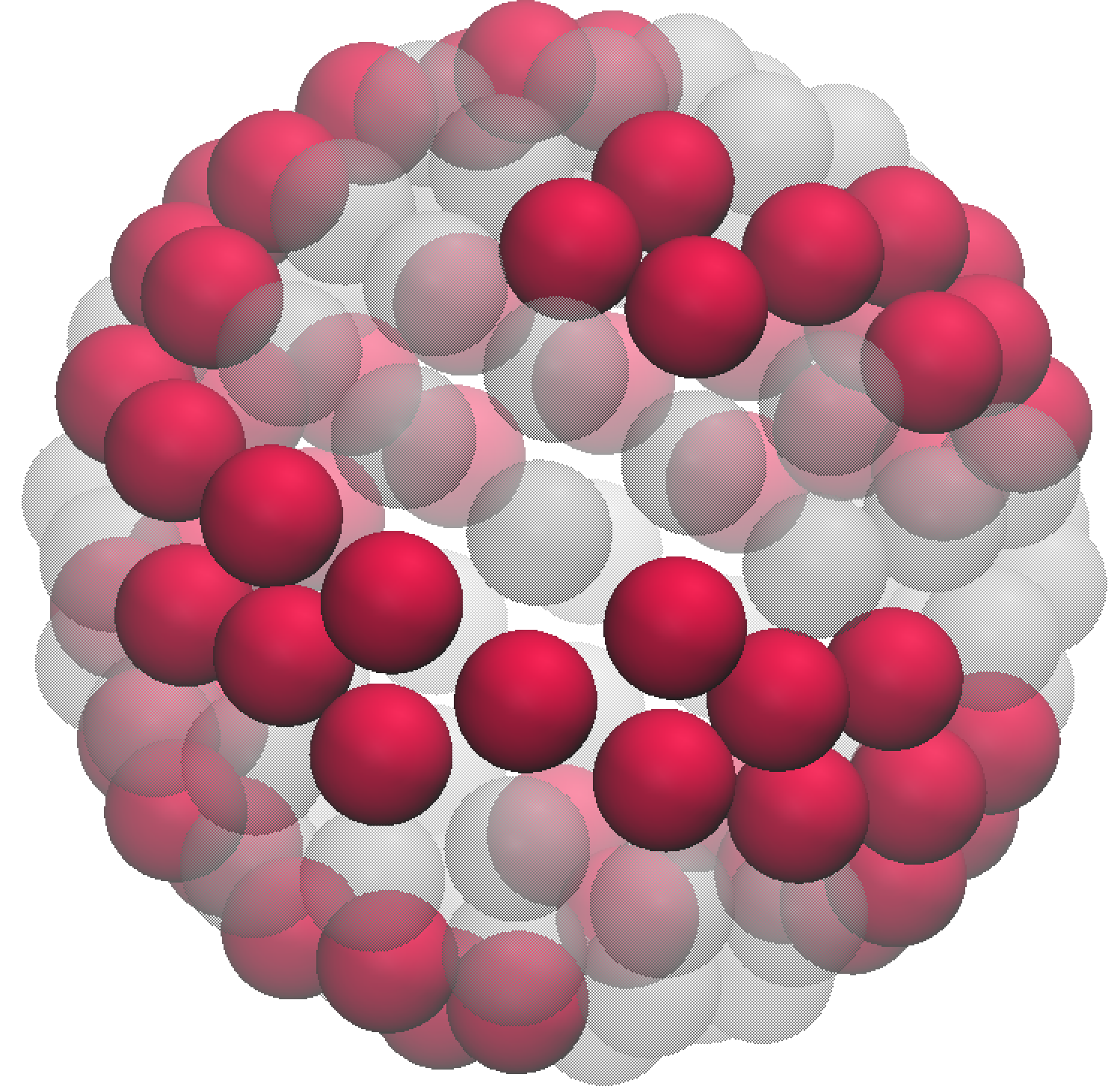}\,\,\,\,\,
\includegraphics[width=3.5cm]{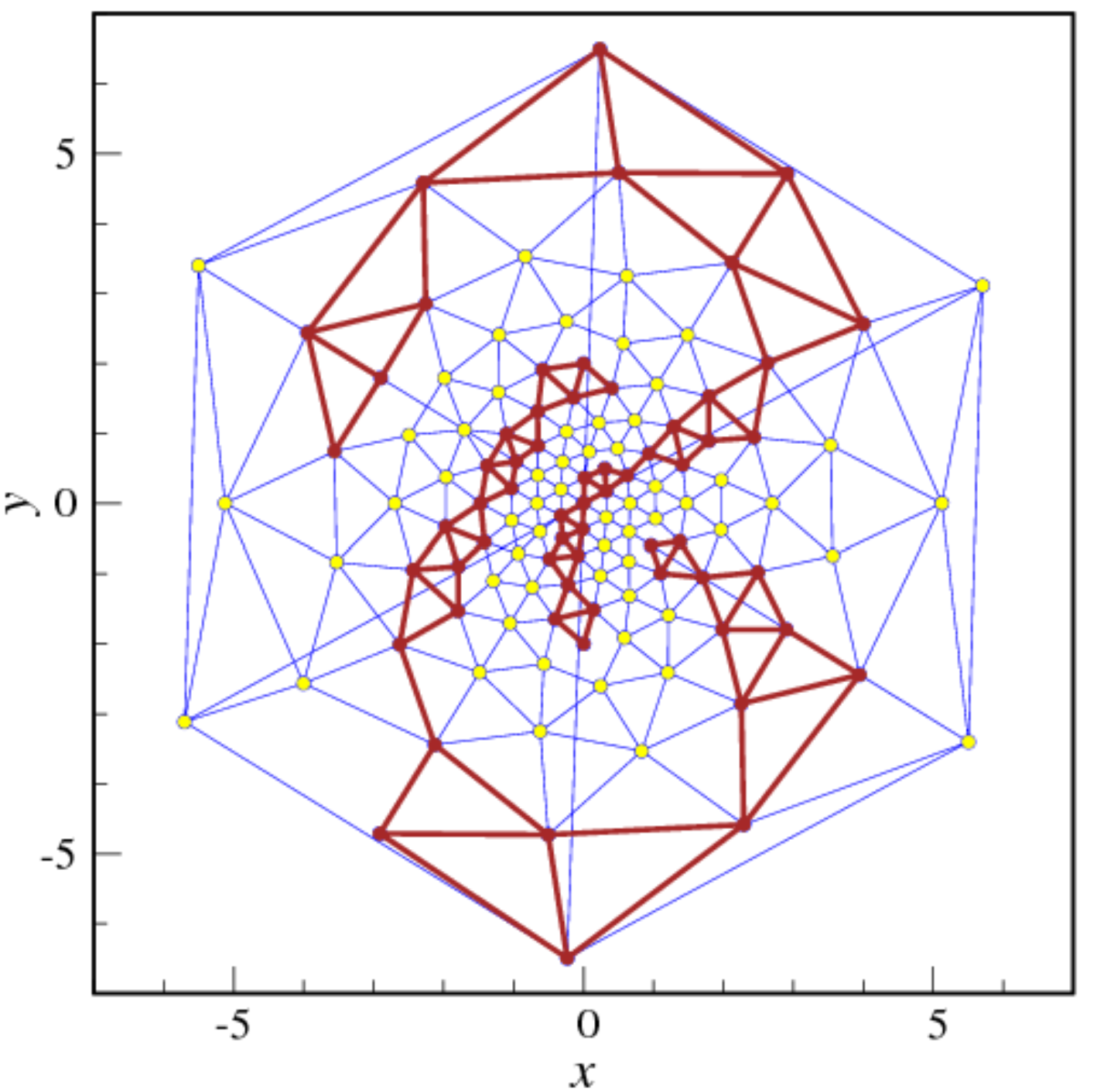}
\\
\includegraphics[width=3.5cm]{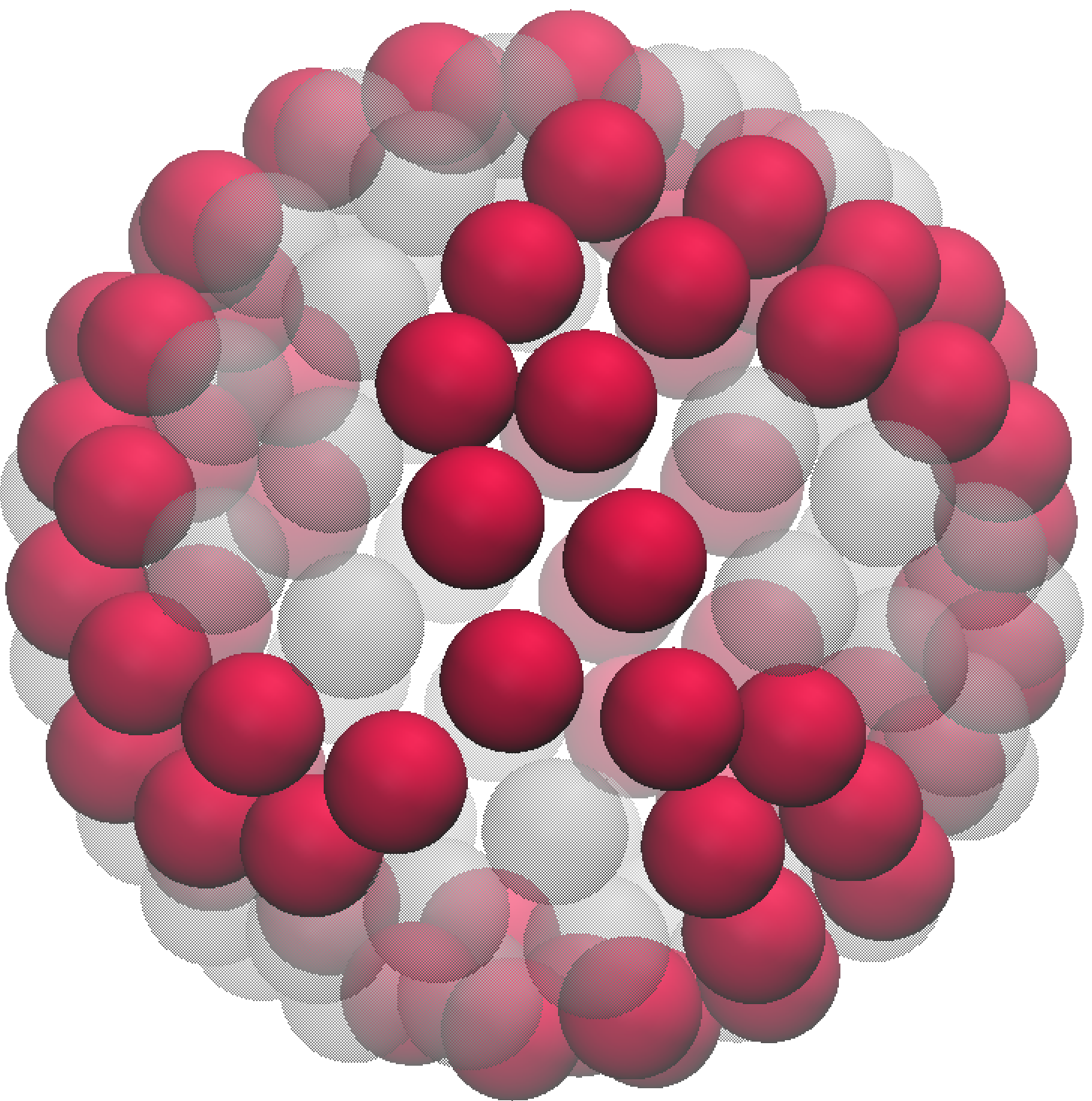}\,\,\,\,\,
\includegraphics[width=3.5cm]{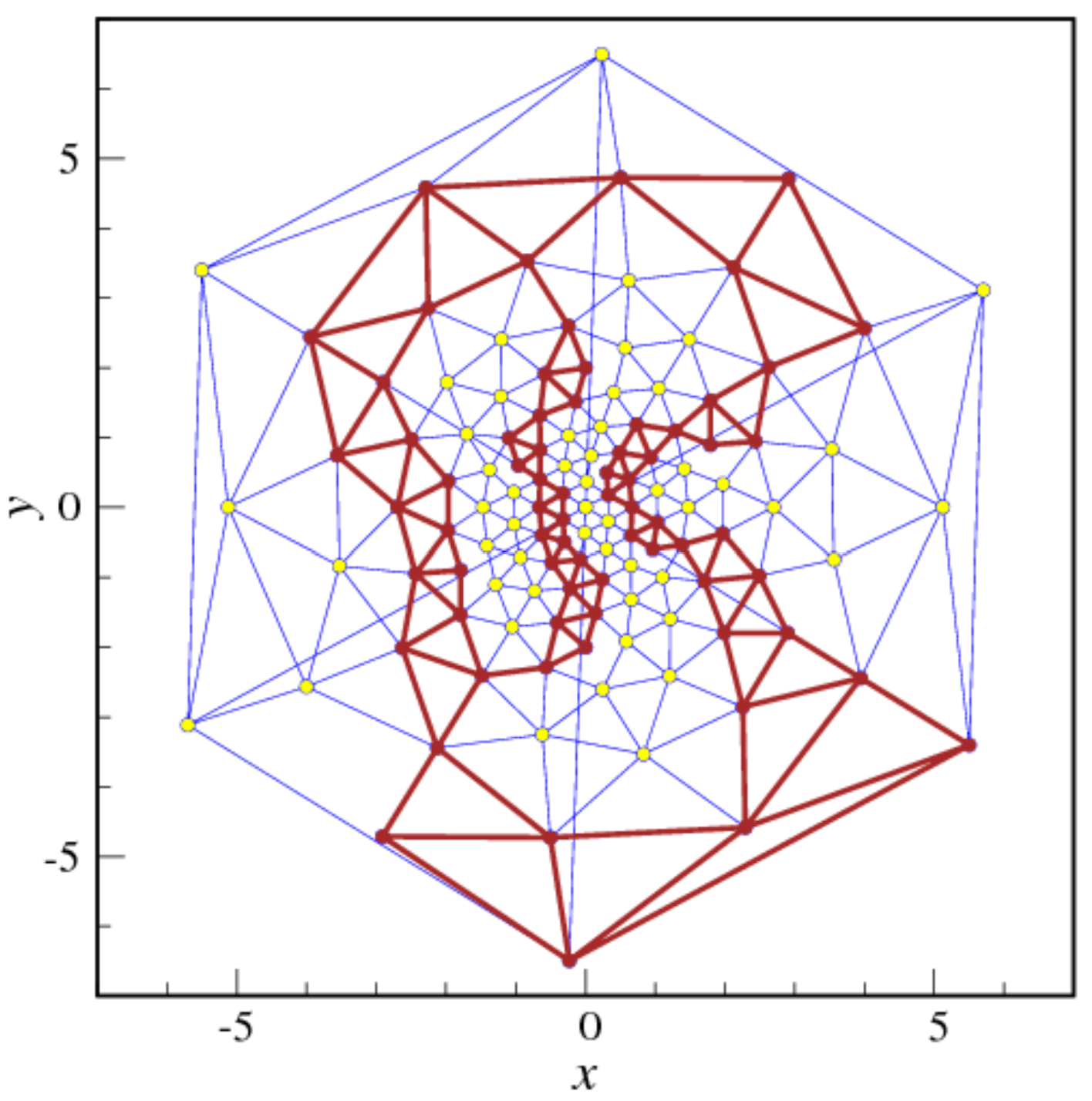}
\caption{HPCD model, $(-1,2,1)$ interaction:
minimum-energy configurations for $N=62$ (left top) and $N=70$ (left bottom), along with their stereographic projections (right).}
\label{hpcd-121}
\end{figure}

\begin{figure*}
\includegraphics[width=4cm]{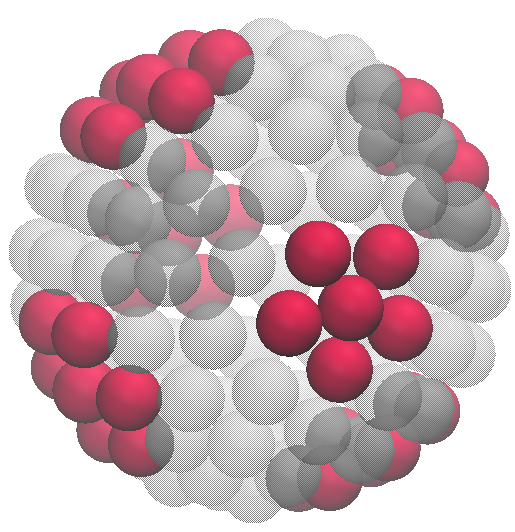}\,\,\,\,\,
\includegraphics[width=4cm]{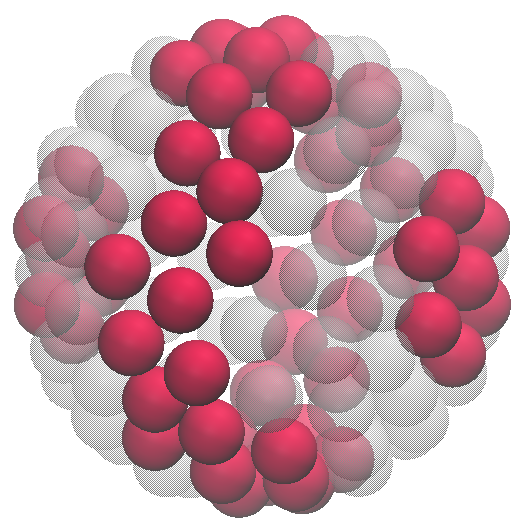}\,\,\,\,\,
\includegraphics[width=4cm]{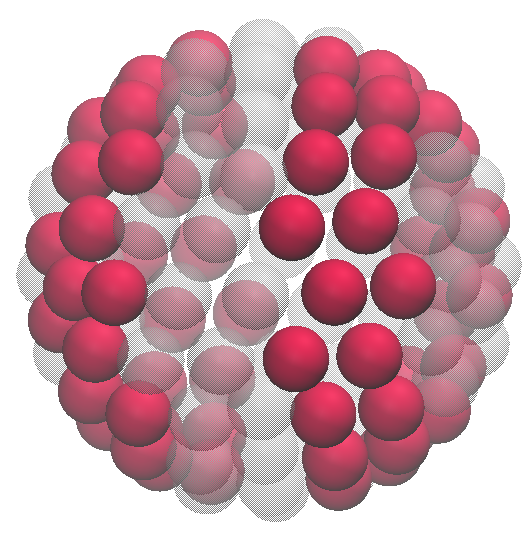}
\caption{HPCD model, $(-1,0,1)$ interaction:
minimum-energy configurations for $N=40,56,64$.}
\label{hpcd-101}
\end{figure*}

As far as Lennard-Jones-like interactions are concerned, for the parameter set $(1,0,-1)$ three distinct phases are observed (see Fig.~\ref{hpcd10-1NES} left).
$N=32$ particles are placed on the vertices of a pentakis dodecahedron.
Judging on the persistence in temperature of the relative plateau in $N(\mu)$, this phase is very robust to thermal fluctuations.
Increasing the chemical potential, a phase transition then occurs at $\mu=0$ into a $N=58$ phase characterized by worm-like structures.
This transition is isoenergetic at $T=0$, whereas at, say, $T=0.5$ the energy of the $N=58$ phase is sensibly higher (Fig.~\ref{hpcd10-1NES} middle).
Eventually, a further transition occurs to a phase with $N=90$, where the system adopts the structure of a rectified truncated icosahedron~\cite{visualpolyhedra}.

As shown by the entropy profile as a function of $\mu$, see Fig.~\ref{hpcd10-1NES} right, each phase transition is accompanied by a peak in the entropy.
This behavior arises from the increased number of microstates available to the system near each phase crossover.

Focusing on the ``transition'' from $N=58$ to $N=90$, occurring at $\mu\approx 0.95$, the generalized grand potential (GGP) $\widetilde{\Omega}_{\mathcal{N,E}}$ is plotted as a function of ${\cal N}$ and ${\cal E}$ in Fig.~\ref{omegatilde}, at low temperature ($T=0.05$).
In particular, we here illustrate the evolution of the two GGP minima across the transition.
At $\mu=0.80$, the global GGP minimum falls at $N=58,E=-90$, and a high free-energy barrier separates this minimum from the relative minimum at $N=90,E=-60$.
At the phase transition, the two minima have exactly the same height, meaning that the increase in chemical potential has canceled the difference between the minima.
At still higher $\mu$, the phase transition has already occurred, since the global GGP minimum is now located at $N=90,E=-60$.
The endurance of the free-energy barrier across the transition indicates that the latter can be classified first-order-like.

When the strength of the first-neighbor repulsion is doubled, we still observe the above mentioned $N=32$ and $N=58$ phases.
However, a new phase appears at $N=62$, where the particles form two interlaced worms.
Furthermore, the rectified truncated icosahedron at $N=90$ is no longer present, being instead replaced by a non-regular polyhedral configuration at $N=92$.
Finally considering Lennard-Jones-like interactions with a minimum at second-neighbor distance, for example $(1,-1,0)$, only two distinct phases are found.
The system organizes itself in either a pentakis icosidodecahedron configuration ($N=42$) or its dual structure, the chamfered dodecahedron~\cite{visualpolyhedra} ($N=80$).
The overall scenario remains unchanged when including a weaker third-neighbor attraction.


\begin{figure*}
\includegraphics[width=5cm]
{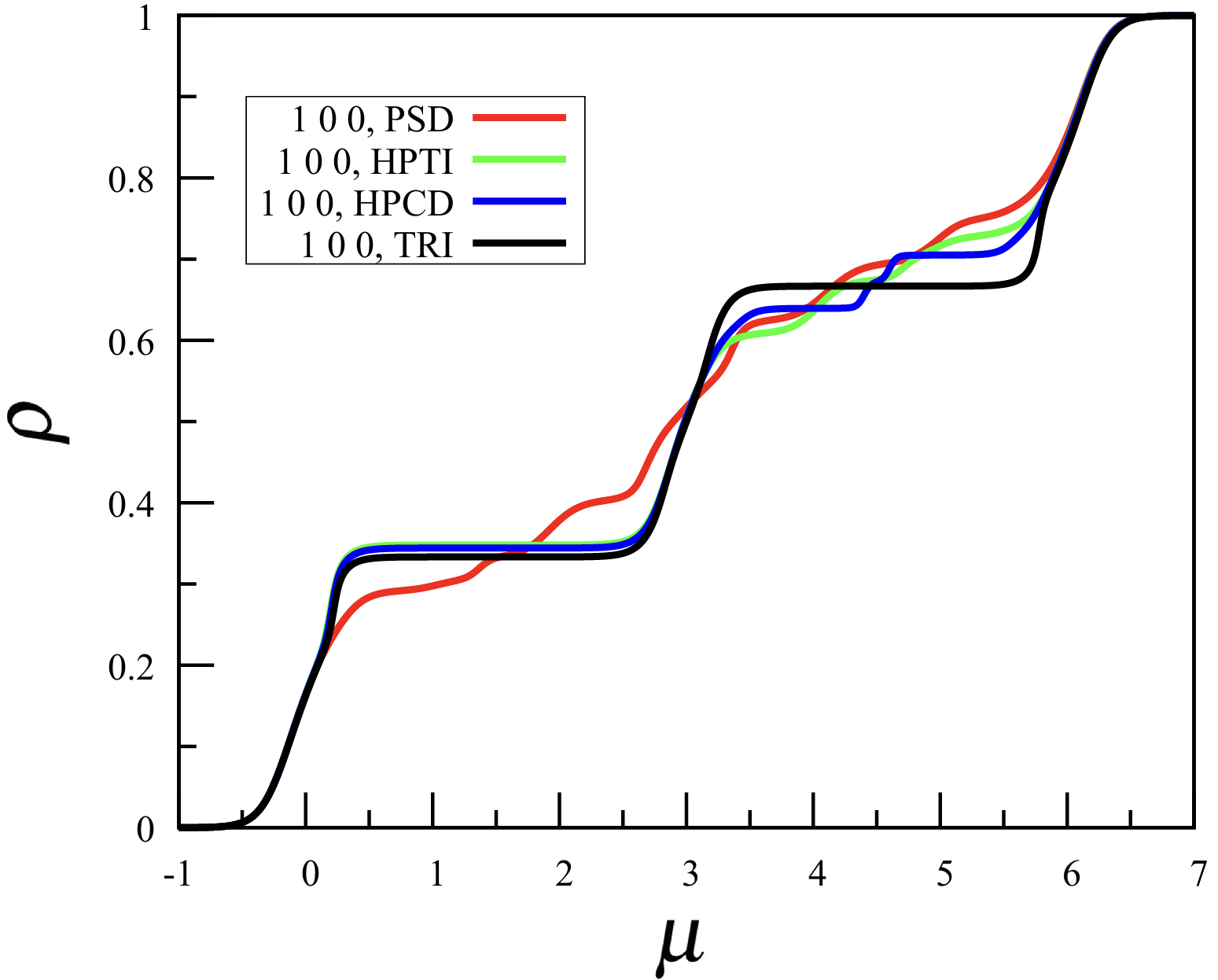}\,\,
\includegraphics[width=5cm]{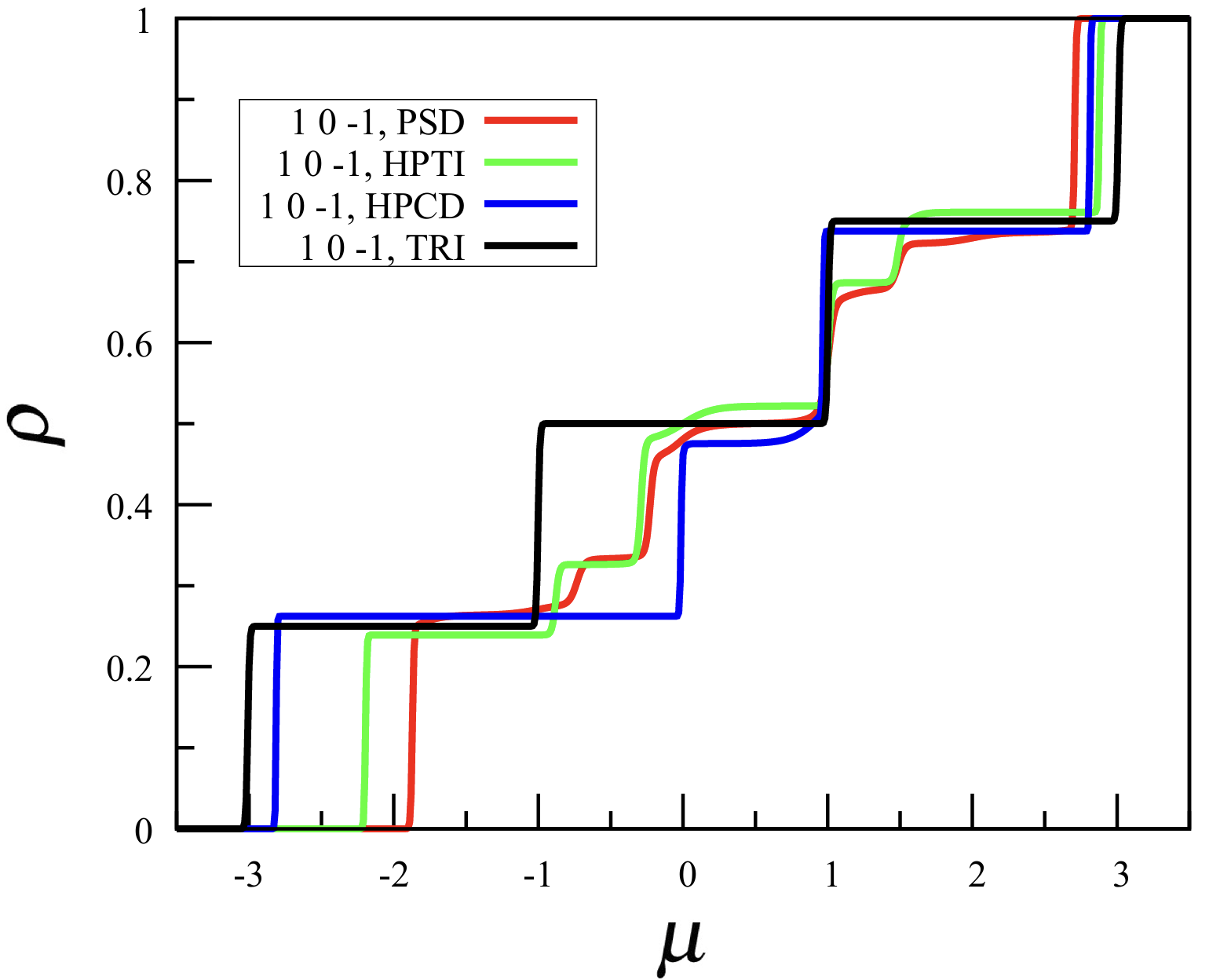}\,\,
\includegraphics[width=5cm]{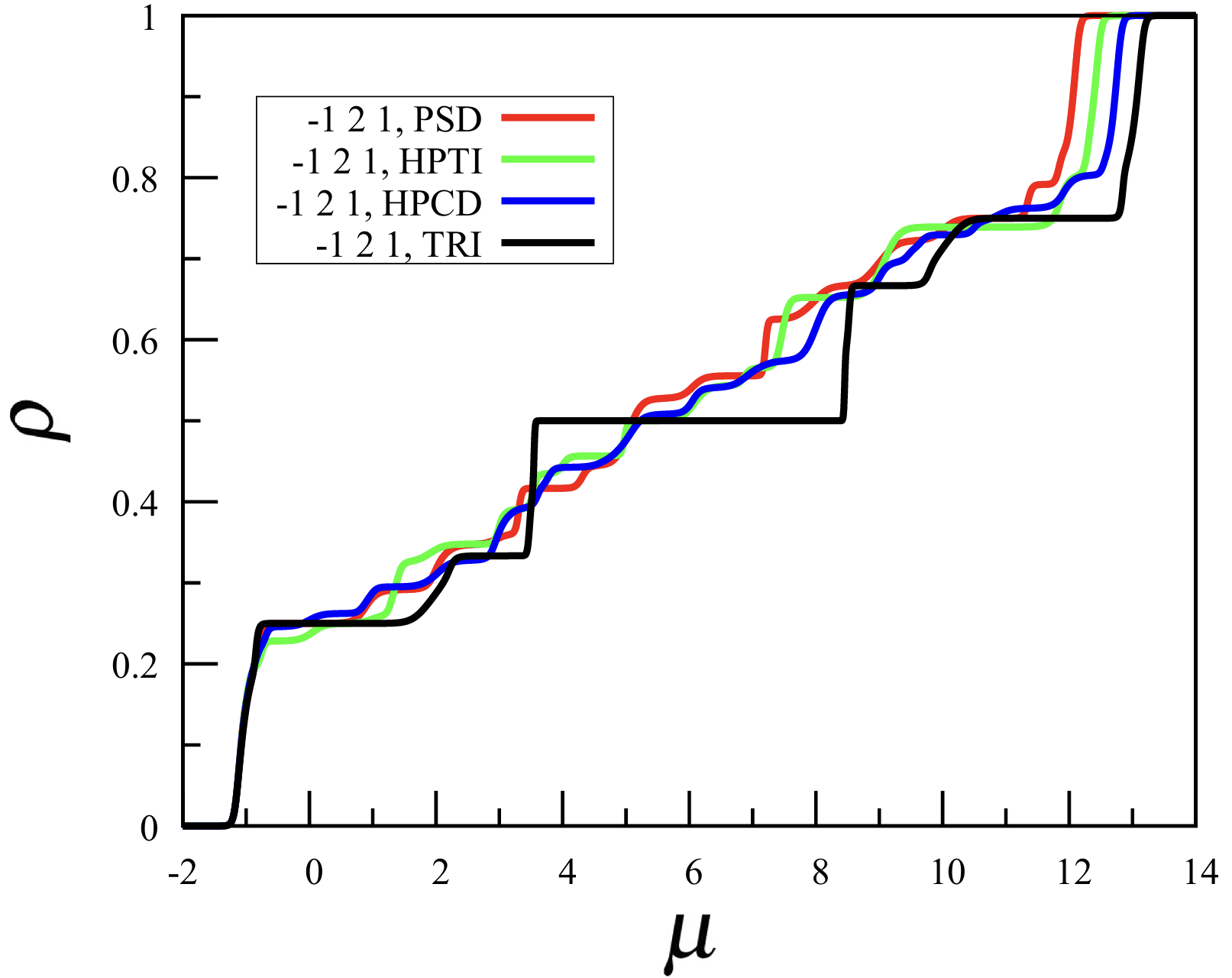}
\caption{$\rho(\mu)$ profile at $T=0.1$ for three geodesic grids and the triangular lattice:
$(1,0,0)$ (left);
$(1,0,-1)$ (middle);  
$(-1,2,1)$ (right).}
\label{confronti}
\end{figure*}

Moving to SALR interactions, the $(-1,2,1)$ set of parameters still offers a wide variety of low-temperature structures.
At relatively low density, the system goes through a sequence of microphases corresponding to cluster-crystal states.
In particular, for $N=30$ ten clusters are arranged at the vertices of a pentagonal antiprism, while for $N=36$ an icosahedron of clusters emerges, similarly to the $(1,1,1)$ case.
As the density increases, the clusters merge, forming six worms comprised of eight particles each ($N=48$).
These worms subsequently join into two longer worms ($N=54-62$, see Fig.~\ref{hpcd-121}), eventually coalescing in a branched configuration with $N=70$ particles.
At high density, the system exhibits cluster-crystal configurations, but with the roles of particles and holes interchanged.
Reducing the height and width of the repulsion ramp, e.g., $(-1,1,0)$, we observe two distinct phases only, for $N=48$ and $N=54$.
In both cases we find a mixture of clusters and worms.
Finally, we analyzed the parameter set $(-1,0,1)$, see Fig.~\ref{hpcd-101}.
At low density we observe pentagon-shaped clusters, while at $N=40$ four hexagonal and two pentagonal clusters sit at the vertices of an octahedron.
As the density increases, we see structures of a type already encountered, i.e., a ring of particles with hexagonal clusters at poles ($N=56$) or two rings ($N=64$).

\subsection{Comparison with the triangular lattice}

\begin{figure}
\includegraphics[width=3.7cm]{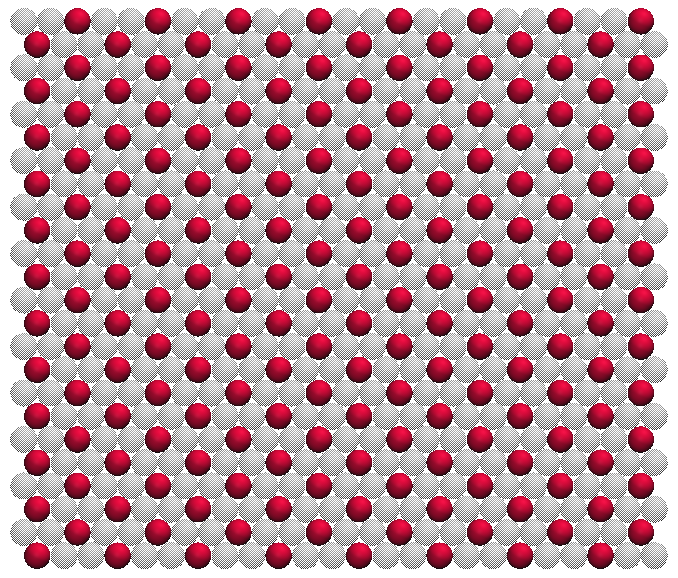}\,\,\,\,\,
\includegraphics[width=3.7cm]{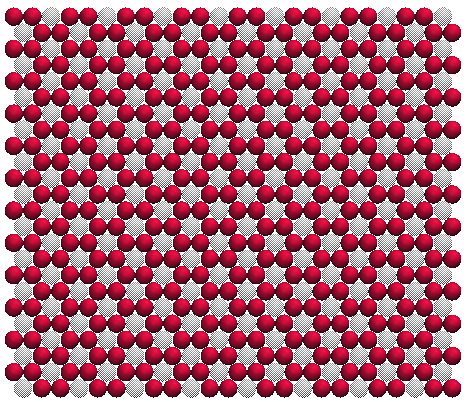}
\caption{Minimum-energy configurations for a triangular-lattice gas with parameters $(1,0,0)$:
$\rho\approx 0.33$ (left) and $\rho\approx 0.67$ (right).}
\label{100lattices}
\end{figure}

\begin{figure*}
\includegraphics[width=4.5cm]{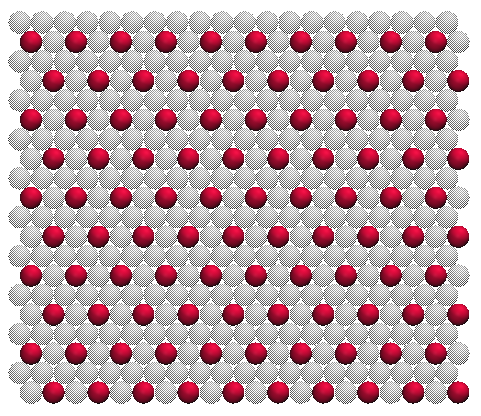}\,\,\,\,\,
\includegraphics[width=4.5cm]{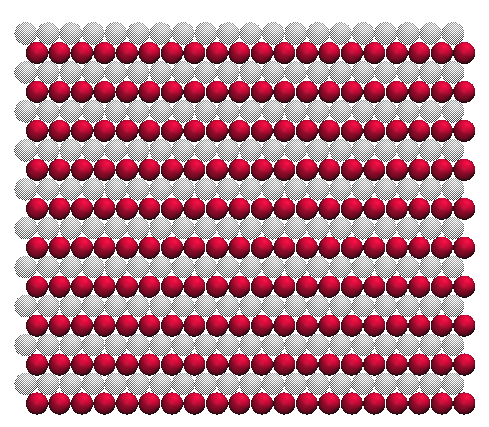}\,\,\,\,\,
\includegraphics[width=4.5cm]{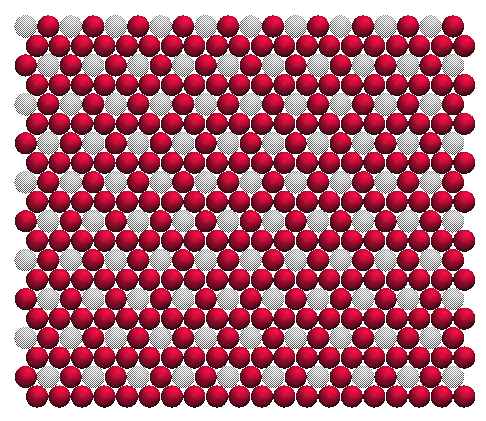}
\caption{Minimum-energy configurations for a triangular-lattice gas with parameters $(1,0,-1)$:
$\rho\approx 0.25$ (left), $\rho\approx 0.50$ (middle), and $\rho\approx 0.75$ (right).}
\label{10-1lattices}
\end{figure*}

\begin{figure*}
\includegraphics[width=4.5cm]{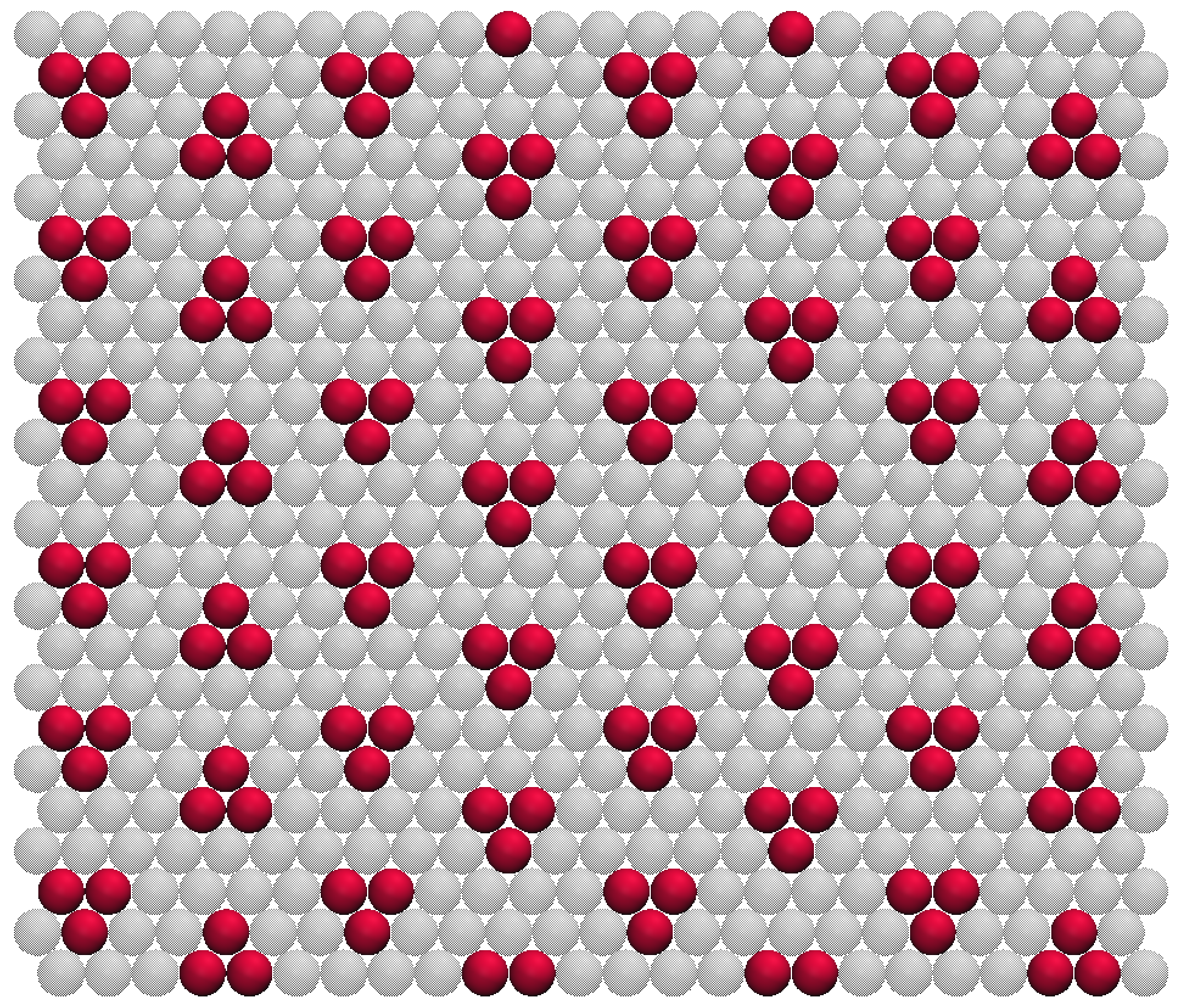}\,\,\,\,\,
\includegraphics[width=4.5cm]{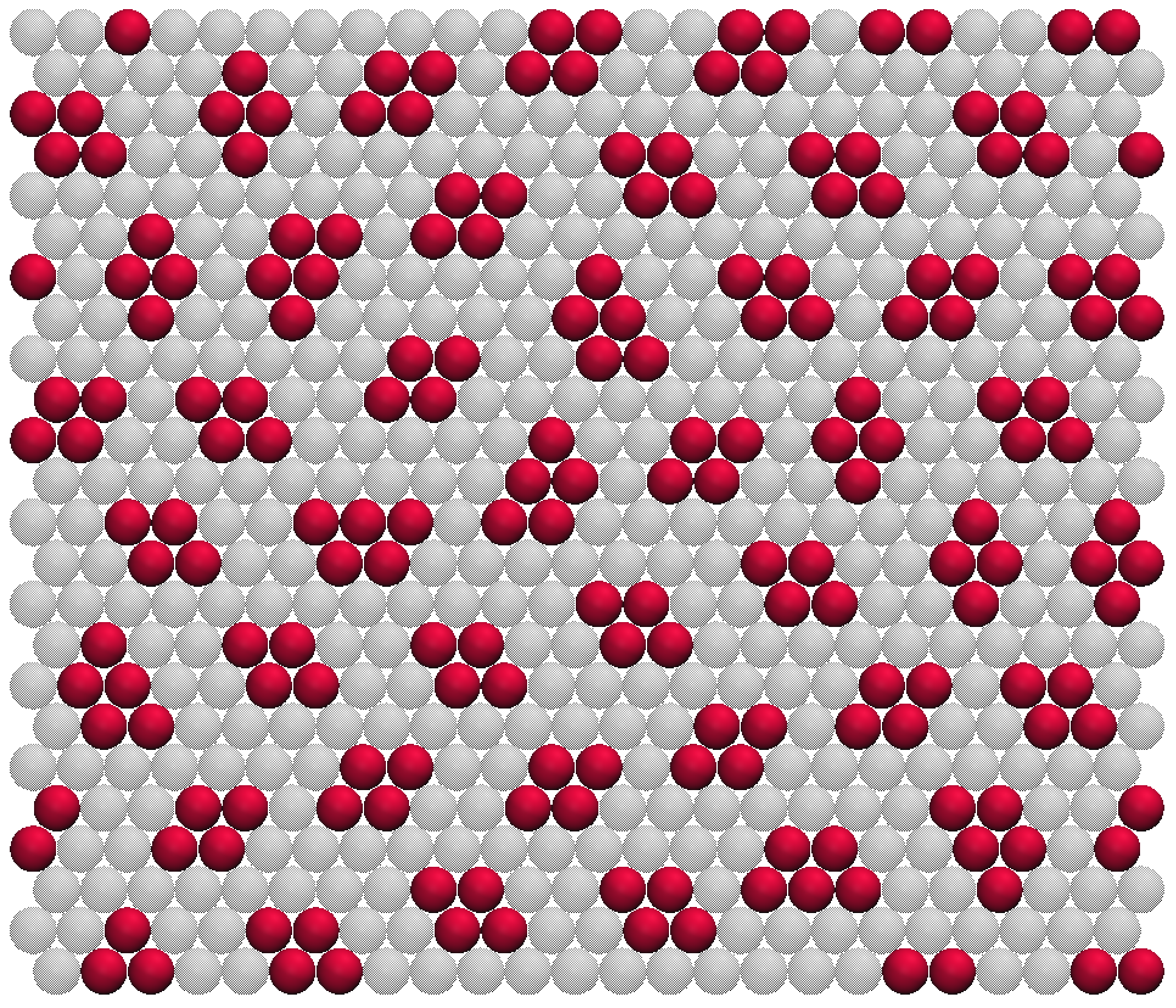}\,\,\,\,\,
\includegraphics[width=4.5cm]{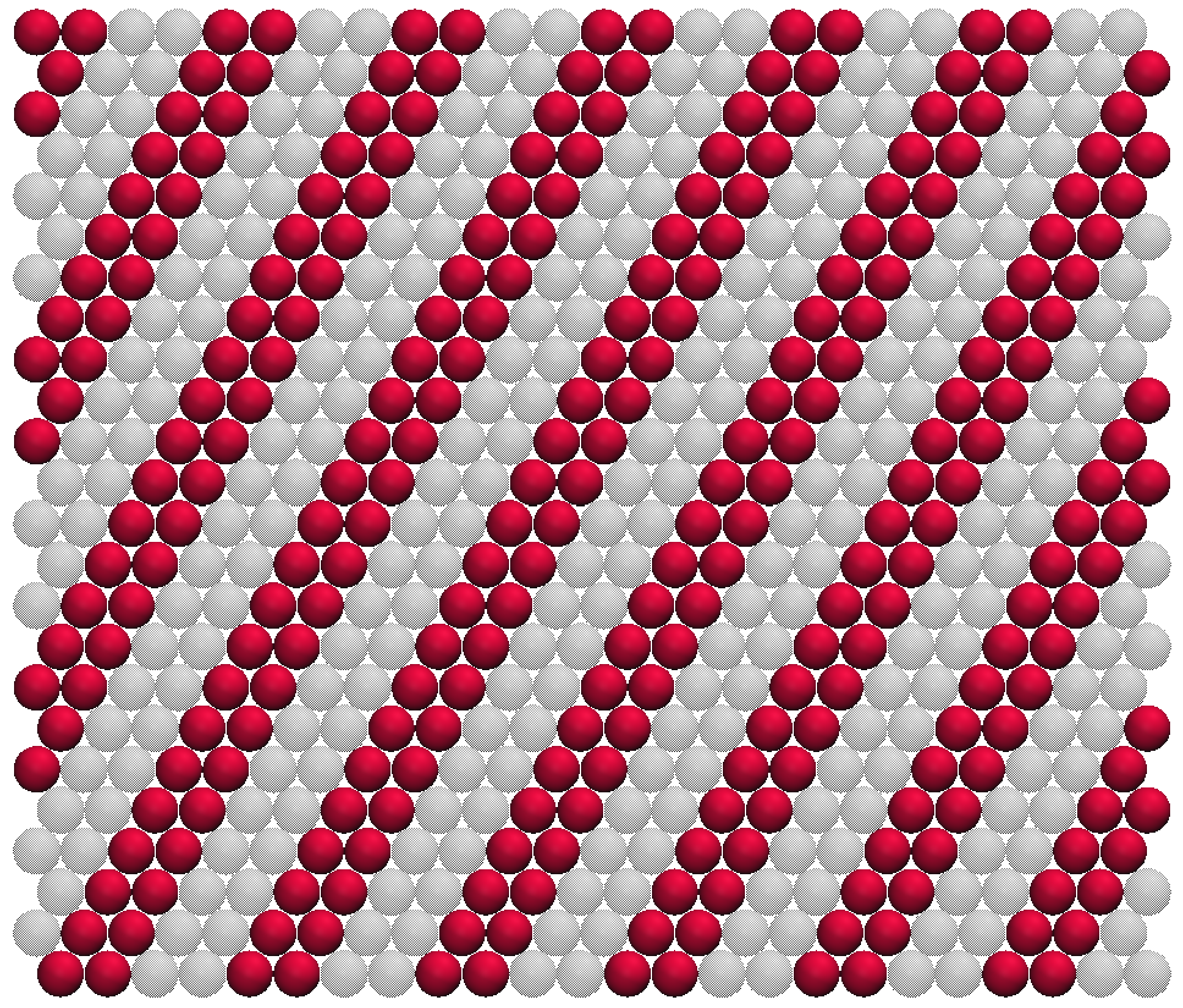}
\caption{Typical configurations obtained via Monte Carlo simulations for a triangular lattice with parameters $(-1,2,1)$, for $T=0.2$:
from left to right, $\rho\approx 0.25,0.33$, and 0.50.
The leftmost and rightmost configurations are also minimum-energy configurations.
The cluster crystal of density 0.25 is not unique, since the orientation of all the triangles in a given column can be inverted at zero cost.}
\label{-121lattices}
\end{figure*}

To understand how a curved space can influence the thermodynamic properties of a lattice-gas system, we compare --- for a few interactions --- the results presented in sections A-C with those relative to the triangular-lattice gas.
We remark that perfect particle-hole symmetry holds for the triangular-lattice gas, irrespective of the interaction (see Appendix A).
For the latter system we still employ the WL method, choosing a $12\times 12$ triangular lattice with periodic boundary conditions; the structures found at low temperature are then confirmed by Metropolis Monte Carlo simulations on larger lattices.
Specifically, we analyze three distinct sets of parameters, each corresponding to a different type of interaction.

First of all, we consider the core-corona $(1,0,0)$ repulsion.
The profile of the density as a function of $\mu$ is plotted in Fig.~\ref{confronti} (left) for $T=0.1$.
The overall behavior of the density on spherical grids is consistent with that found on the triangular lattice.
Notably, the latter profile shows two distinct plateaus at $\rho\approx 0.33$ and $\rho\approx 0.67$, corresponding to the structures depicted in Fig.~\ref{100lattices}.
Of the various curves in Fig.~\ref{confronti}, the density curve on the biggest grid (HPCD) is the most similar to that on the triangular lattice.

For the Lennard-Jones-like $(1,0,-1)$ interaction, the density profile $\rho(\mu)$ is reported in Fig.~\ref{confronti} (middle), again for $T=0.1$.
As seen in the picture, a direct comparison between the various curves is challenging.
However, it is fair to say that the HPCD grid, which has the smallest curvature among all grids, provides the best approximation to the triangular lattice.
On this lattice, $\rho(\mu)$ is characterized by three distinct plateaus at $0.25,0.50$, and 0.75.
The configurations corresponding to these densities are illustrated in Fig.~\ref{10-1lattices}.
At $\rho=0.25$ the particles form a triangular crystal whose lattice spacing is twice larger than that of the underlying lattice;
at $\rho=0.75$ occupied and empty sites are interchanged with respect to $\rho=0.25$.
Instead, at $\rho=0.50$ particles are lined up to form parallel stripes.
These configurations have an obvious relationship with those observed on geodesic grids at low $T$.
Indeed, at low and high densities curved systems typically exhibit polyhedral structures, whereas at intermediate densities they are characterized by worm-like structures.

Finally, we consider the SALR interaction $(-1,2,1)$.
The various density profiles for $T=0.1$ are compared in Fig.~\ref{confronti} (right).
At this temperature, the density curve of the triangular-lattice gas exhibits five plateaus; the configurations at $\rho\approx 0.25$ and 0.50 are identical to those reported in Fig.~\ref{-121lattices}, which are outcomes of a simulation carried out at $T=0.2$.
For $\rho\approx 0.33$ the minimum-energy configuration in a $12\times 12$ lattice consists of a crystal of four-particle clusters;
in the simulation, which uses a larger sample, most (but not all) of the clusters are of the same sort.

Although the comparison between density profiles is not particularly valuable due to strong curvature effects, cluster crystals exist in both geometries.
Similarly, the worm-like structures observed on spherical grids near $\rho=0.50$ show a characteristic width of two particles, in reasonable agreement with the stripes found on the triangular lattice.
In closing, it is worth noting that the $\rho(\mu)$ curve for the SALR system with parameters $(-1,1,0)$ shows a single plateau at $\rho\approx 0.50$, corresponding to two-row stripes.
This agrees with the view that cluster crystals require a sufficiently high and long repulsive ramp to be stabilized~\cite{zhuang2016recent}.

\section{Conclusions}

Using WL sampling, we have worked out the exact phase behavior of a variety of lattice-gas systems defined on geodesic grids.
Among spherical grids, geodesic grids provide the closest approximations to the triangular lattice.
Originally, the aim of this analysis was twofold:
firstly, to provide a thorough classification of low-temperature ``phases'', so as to draw a correspondence between particle interaction and global system structure.
On the other hand, our study had also the purpose to assess the effects of geometric frustration on the ground states of the triangular lattice gas, in a working framework where the amount of curvature can be tuned.

We have considered three types of pair interactions, extending up to third neighbors:
core-corona, Lennard-Jones-like, and SALR interactions.
In the former case, a common thread among the various grids is the existence of ground states with polyhedral order at low density and worm-like arrangements at moderate densities.
For interactions of Lennard-Jones type, structures with polyhedral order are only observed when the minimum of the interaction energy falls at second- or third-neighbor distance.
In the latter case we also find worm-like structures at moderate density.
Finally, when couplings are adjusted to a SALR interaction, the only ordered structures at low density are cluster crystals, while a multitude of worm-like configurations are stabilized at intermediate densities.

Interestingly, our survey reveals that striped patterns exist as lowest-energy configurations for all kinds of interactions.
As far as we know, this finding is novel for Lennard-Jones type of interactions in which the attraction encompasses third-, rather than first- or second-neighbor particles on the lattice.

The structural complexity of lattice gases on spherical grids gets lost when moving to the triangular lattice, despite an evident relationship remaining between the structure of the lowest-energy configurations in flat space and the recurrent motifs on curved grids.
As the curvature is gradually reduced, the profile of the density as a function of chemical potential shows a clear trend towards the density curve for the triangular-lattice gas.
The relationship between the triangular lattice and geodesic grids is similar to the approximation of planar lattices by sequences of aperiodic tilings~\cite{Matsubara2024}.
The arrangements observed on spherical grids can also be compared against those exhibited by particles confined on a spherical surface.
For example, at low temperature a SALR potential consisting of the sum of Lennard-Jones and Yukawa potentials promotes the formation of striped patterns on a sphere~\cite{Pekalski2018} that are essentially identical to those found on the largest geodesic grids.

To see how our results could turn useful in practice, suppose that we want to functionalize spherical colloidal microparticles with attractive patches of desired number and shape.
The patches are made of smaller particles deposited on the surface of the microparticle;
by employing suitably interacting smaller particles, and choosing their size appropriately, we can in principle bring them to self-assemble in the form of patches, somehow like illustrated in Fig.~\ref{hpti-121} or \ref{hpcd111}.
A recent method by which the ``valence'' of a DNA meshframe (produced with the technique of DNA origami~\cite{Benson2015}) can be programmed to coordinate nanoparticles into pre-defined cluster architectures is described in Ref.~\cite{Sun2020}.
Interestingly, a specific example made in the latter paper is a pentakis-icosidodecahedral mesh.
Clearly, if we knew in advance the shape of the interaction giving rise to specific self-assembled structures, we could gain a better control on the synthesis of patchy particles.
Our paper aims to be a contribution in this direction.

Among future prospects, we plan to extend the study of lattice gases on spherical grids to binary mixtures, especially with the aim of establishing conditions for the appearance of stripes and other ordered patterns at high density~\cite{Prestipino2023}.

\appendix*
\section{Proof of a symmetry property}

The $N(\mu)$ curves in Fig.~\ref{symmetries} are manifestly symmetric around $N=M/2=46$;
we have also found the same particle-hole symmetry in other cases not commented in the text.
Below, we provide an explanation for this fact.

In order for the mentioned symmetry to hold, there must exist a chemical potential $\mu_0$ such that
\begin{equation}
N(\mu_0+\Delta\mu)=M-N(\mu_0-\Delta\mu)\,.
\label{eqa1}
\end{equation}
By working out the rhs of Eq.~\ref{eqa1}, we obtain:
\begin{eqnarray}
M-N(\mu)&=&\langle M-{\cal N}\rangle(\mu)
\nonumber \\
&=&\frac{\sum_{\{c\}}\left(M-\sum_ic_i\right)e^{\beta\mu\sum_ic_i-\beta H[c]}}{\sum_{\{c\}} e^{\beta\mu\sum_ic_i-\beta H[c]}}
\nonumber \\
&=&\frac{\sum_{\{c\}}\left(\sum_i(1-c_i)\right)e^{\beta\mu\sum_ic_i-\beta H[c]}}{\sum_{\{c\}} e^{\beta\mu\sum_ic_i-\beta H[c]}}\,.
\nonumber \\
\label{eqa2}
\end{eqnarray}
Upon substituting $c_i\rightarrow 1-c_i$ for all $i$, we arrive at
\begin{equation}
M-N(\mu)=\frac{\sum_{\{c\}}\left(\sum_ic_i\right)e^{\beta\mu\sum_i(1-c_i)-\beta H[1-c]}}{\sum_{\{c\}} e^{\beta\mu\sum_i(1-c_i)-\beta H[1-c]}}\,.
\label{eqa3}
\end{equation}
Now, the calculation of $H[1-c]$ yields:
\begin{eqnarray}
H[1-c]&=&u_1\sum_{\rm 1NP}(1-c_i)(1-c_j)+\ldots
\nonumber \\
&=&u_1\left(\sum_{\rm 1NP}1-2\sum_{\rm 1NP}c_i+\sum_{\rm 1NP}c_ic_j\right)+\ldots\,,
\nonumber \\
\label{eqa4}
\end{eqnarray}
omitting similar terms containing $u_2$ and $u_3$.
First, we have:
\begin{eqnarray}
\sum_{\rm 1NP}1=\sum_{\rm 2NP}1=\frac{1}{2}(12\cdot 5+60\cdot 6+20\cdot 6)=270\,;
\nonumber \\
\sum_{\rm 3NP}1=\frac{1}{2}(12\cdot 5+60\cdot 5+20\cdot 6)=240\,.
\nonumber \\
\label{eqa5}
\end{eqnarray}
The evaluation of $\sum_{\rm 1NP}c_i$ is more intricate.
Denoting the nearest neighbors of site $i$ as NN$_{i}$, the fivefold sites as ${\rm 5s}$, and the two sets of sixfold sites (see Table \ref{table1}) as ${\rm 6s_1}$ and ${\rm 6s_2}$, we obtain:
\begin{eqnarray}
\sum_{\rm 1NP}c_i&=&\frac{1}{2}\sum_ic_i\sum_{j\in{\rm NN}_i}1
\nonumber \\
&=&\frac{1}{2}\left(\sum_{\rm 5s}c_i\cdot 5+\sum_{\rm 6s_1}c_i\cdot 6+\sum_{\rm 6s_2}c_i\cdot 6\right)
\nonumber \\
&=&\frac{5}{2}\sum_ic_i+\frac{1}{2}\sum_{\rm 6s_1}c_i+\frac{1}{2}\sum_{\rm 6s_2}c_i\,.
\label{eqa6}
\end{eqnarray}
Similarly,
\begin{eqnarray}
\sum_{\rm 2NP}c_i&=&\frac{1}{2}\left(\sum_{\rm 5s}c_i\cdot 5+\sum_{\rm 6s_1}c_i\cdot 6+\sum_{\rm 6s_2}c_i\cdot 6\right)
\nonumber \\
&=&\frac{5}{2}\sum_ic_i+\frac{1}{2}\sum_{\rm 6s_1}c_i+\frac{1}{2}\sum_{\rm 6s_2}c_i
\label{eqa7}
\end{eqnarray}
and
\begin{eqnarray}
\sum_{\rm 3NP}c_i&=&\frac{1}{2}\left(\sum_{\rm 5s}c_i\cdot 5+\sum_{\rm 6s_1}c_i\cdot 5+\sum_{\rm 6s_2}c_i\cdot 6\right)
\nonumber \\
&=&\frac{5}2\sum_ic_i+\frac{1}{2}\sum_{\rm 6s_2}c_i\,.
\label{eqa8}
\end{eqnarray}
The last terms in Eqs.~(\ref{eqa6})-(\ref{eqa8}) are configuration-dependent;
however, if $u_1+u_2=0$ and $u_3=0$, then the contribution from those terms vanish and we simply get:
\begin{equation}
H[1-c]=H[c]\,.
\label{eqa9}
\end{equation}
Eq.~\ref{eqa3} then becomes:
\begin{eqnarray}
M-N(\mu)&=&\frac{\sum_{\{c\}}\left(\sum_ic_i\right)e^{-\beta\mu\sum_ic_i}e^{-\beta H[c]}}{\sum_{\{c\}} e^{-\beta\mu\sum_ic_i}e^{-\beta H[c]}}
\nonumber \\
&=&N(-\mu)\,.
\label{eqa10}
\end{eqnarray}
Upon comparing Eqs.\,\ref{eqa1} and \ref{eqa10}, we immediately see that $\mu_0=0$ and $\Delta\mu=\mu$.

By performing an analogous derivation for the HPCD model, we see that the symmetry condition remains unchanged, implying that $u_1+u_2=0$ and $u_3=0$ hold for both models.
In contrast, for the PSD model no special condition on $u_3$ is required.
As a result, the symmetry condition simplifies to $u_1+u_2=0$.
Under this assumption, for the PSD model we obtain:
\begin{equation}
\mu_0=\frac{5}{2}u_3
\,\,\,\,\,\,{\rm and}\,\,\,\,\,\,
\Delta\mu=\frac{5}{2}u_3-\mu\,.
\label{eqa11}
\end{equation}

Looking at Fig.~\ref{confronti}, it appears that the density profile of the triangular-lattice gas is always symmetric, regardless of the energy couplings.
The same figure indicates that the abscissa $\mu_0$ of the inversion-symmetry center is instead interaction-dependent.
Following the same steps of the above derivation, for a triangular lattice of $M$ sites we arrive at:
\begin{equation}
H[1-c]=3M(u_1+u_2+u_3)-6(u_1+u_2+u_3)\sum_i c_i+H[c]\,.
\label{eqa12}
\end{equation}
At variance with the previous cases, no condition is to be imposed to obtain Eq.~(\ref{eqa12}).
Thus, for any choice of the energy couplings we have:
\begin{equation}
M-N(\mu)=N(6(u_1+u_2+u_3)-\mu)\,,
\label{eqa13}
\end{equation}
finally yielding
\begin{equation}
\mu_0=3(u_1+u_2+u_3)\,.
\label{eqa14}
\end{equation}
It is easy to verify that Eq.~\ref{eqa14} is consistent with the curves plotted in Fig.~\ref{confronti}.
\bibliography{paper}
\end{document}